\documentclass[lettersize,journal]{IEEEtran}

\usepackage{amsmath,amsfonts}
\usepackage{algorithmic}
\usepackage{array}
\usepackage[caption=false,font=normalsize,labelfont=sf,textfont=sf]{subfig}
\usepackage{textcomp}
\usepackage{stfloats}
\usepackage{url}
\usepackage{verbatim}
\usepackage{graphicx}
\def\BibTeX{{\rm B\kern-.05em{\sc i\kern-.025em b}\kern-.08em
    T\kern-.1667em\lower.7ex\hbox{E}\kern-.125emX}}
\usepackage{balance}
\usepackage{color}
\usepackage{xcolor}
\usepackage{colortbl}
\usepackage{tcolorbox}
\usepackage{listings}
\usepackage{makecell}
\newcommand{\tabincell}[2]{\begin{tabular}{@{}#1@{}}#2\end{tabular}} 
\definecolor{ashgrey}{rgb}{0.98, 0.91, 0.71}
\definecolor{grey}{rgb}{0.9, 0.9, 0.9}

\usepackage{multirow}
\usepackage{amsmath,amsfonts}

\usepackage{amssymb}
\usepackage{hyperref} 
\usepackage{booktabs}
\usepackage[utf8]{inputenc}
\usepackage{color}
\usepackage{calligra}
\usepackage{multirow}
\usepackage[normalem]{ulem}
\usepackage{indentfirst}
\usepackage{balance}
\usepackage{url}
\usepackage{enumitem}
\usepackage{nicematrix}
\usepackage{makecell}
\usepackage{fontawesome5}
\usepackage[]{collab}

\collabAuthor{yt}{red}{YT'Note}
\collabAuthor{yc}{blue}{YC'Note}
\collabAuthor{pj}{purple}{PJ'Note}
\collabAuthor{yx}{teal}{Yuxin'Note}
\collabAuthor{zh}{yellow}{ZH'Note}
\collabAuthor{ry}{green}{RY'Note}

\definecolor{codegreen}{rgb}{0,0.6,0}
\definecolor{codegray}{rgb}{0.5,0.5,0.5}
\definecolor{codepurple}{rgb}{0.58,0,0.82}
\definecolor{backcolour}{rgb}{0.95,0.95,0.92}
\definecolor{cerulean}{rgb}{0.0, 0.48, 0.65}
\definecolor{ceruleanblue}{rgb}{0.16, 0.32, 0.75}
\definecolor{cadmiumred}{rgb}{0.89, 0.0, 0.13}

\definecolor{viol}{RGB}{134,0,175}
\definecolor{githubgreen}{RGB}{204, 255, 204}
\definecolor{githubred}{RGB}{255, 224, 224}

\DeclareMathAlphabet{\mathcalligra}{T1}{calligra}{m}{n}

\def\Odata{LogBench-O\xspace}
\def\Tdata{LogBench-T\xspace}
\newcommand{\fixedwidth}[1]{{\ttfamily \small #1}}

\begin{document}
\title{Exploring the Effectiveness of LLMs in Automated Logging Generation: An Empirical Study}

\author{
    \large Yichen Li*\IEEEauthorrefmark{2},
    Yintong Huo*\IEEEauthorrefmark{2}\thanks{* Co-first authors.},
    Zhihan Jiang\IEEEauthorrefmark{2}, 
    Renyi Zhong\IEEEauthorrefmark{2},
    Pinjia He\IEEEauthorrefmark{4},
    Yuxin Su\IEEEauthorrefmark{3},\\ 
    Lionel Briand\IEEEauthorrefmark{5}, and
    Michael R. Lyu\IEEEauthorrefmark{2}\\
  \IEEEauthorblockA{
  \IEEEauthorrefmark{2}The Chinese University of Hong Kong, Hong Kong, China,\\ \{ycli21, ythuo, zhjiang22, ryzhong22, lyu\}@cse.cuhk.edu.hk\\
\IEEEauthorrefmark{3}School of Software Engineering, Sun Yat-sen University, Zhuhai, China, suyx35@mail.sysu.edu.cn\\
  \IEEEauthorrefmark{4}The Chinese University of Hong Kong, Shenzhen, China, hepinjia@cuhk.edu.cn\\
  \IEEEauthorrefmark{5}University of Ottawa, Canada, and the Lero SFI Centre for Software Research, University of Limerick, Ireland, lbriand@uottawa.ca
}
}

\maketitle

\begin{abstract}
Automated logging statement generation supports developers in documenting critical software runtime behavior.
While substantial recent research has focused on retrieval-based and learning-based methods, results suggest they fail to provide appropriate logging statements in real-world complex software.
Given the great success in natural language generation and programming language comprehension, large language models (LLMs) might help developers generate logging statements, but this has not yet been investigated.

To fill the gap, this paper performs the first study on exploring LLMs for logging statement generation.
We first build a logging statement generation dataset, \textit{LogBench}, with two parts: (1) \textit{LogBench-O}: \textit{3,870} methods with \textit{6,849} logging statements collected from GitHub repositories, and (2) \textit{LogBench-T}: the transformed unseen code from \Odata.
Then, we leverage LogBench to evaluate the \textit{effectiveness} and \textit{generalization capabilities} (using \textit{LogBench-T}) of eleven top-performing LLMs, from 60M to 175B parameters. In addition, we examine the performance of these LLMs against classical retrieval-based and machine learning-based logging methods from the era preceding LLMs.
Specifically, we evaluate the logging effectiveness of LLMs by studying their ability to determine logging ingredients and the impact of prompts and external program information.
We further evaluate LLM's logging generalization capabilities using unseen data ({LogBench-T}) derived from code transformation techniques.

While existing LLMs deliver decent predictions on logging levels and logging variables, our study indicates that they only achieve a maximum BLEU score of \textit{0.249}, thus calling for improvements. 
The paper also highlights the importance of prompt constructions and external factors (e.g., programming contexts and code comments) for LLMs' logging performance. 
In addition, we observed that existing LLMs show a significant performance drop (\textit{8.2\%-16.2\%} decrease) when dealing with logging unseen code, revealing their unsatisfactory generalization capabilities.
Based on these findings, we identify five implications and provide practical advice for future logging research.
Our empirical analysis discloses the limitations of current logging approaches while showcasing the potential of LLM-based logging tools, and provides actionable guidance for building more practical models.

\end{abstract}

\vspace{0.4in}

\section{Introduction}

\IEEEPARstart{W}{riting} appropriate logging statements in code is critical for documenting program runtime behavior, supporting various software development tasks. 
Effective logging statements can facilitate performance analysis~\cite{chen2019improving, xu2009detecting} and provide insights for failure identification~\cite{huo2021semparser, huo2023evlog, liu2023scalable,khan2023impact}.
As shown in the example below, a logging statement typically consists of three \textit{ingredients}: a logging level, logging variables, and logging texts~\cite{he2021survey}.
Specifically, as illustrated in the example below, logging level (e.g., \textit{warn}) indicates the severity of a log event; logging variables (e.g., \textit{url}) contain essential run-time information from system states; and logging texts (e.g., \textit{Failed to connect to host: $<>$}) provides a description of the system's activities.
\begin{center}
\small
    \fbox
    {\shortstack[l]{
    \fixedwidth{log.warn("Failed to connect to host: \{\}", url)}
    }}
\end{center}
To help software developers decide the contents of logging statements (i.e., \textit{what-to-log}), logging statement generation tools are built to automatically suggest logging statements given code snippets. Conventional logging suggestion studies~\cite{gholamian2021leveraging, yuan2012characterizing} reveal that similar code tends to have similar logging statements, and thus, a retrieval-based approach is used to suggest similar logging statements from a historical code base~\cite{he2018characterizing}.
However, such retrieval-based approaches are limited to the logging statements encountered in that code base.
% and cannot learn from new data or adapt to changing program behaviors.
% flexibility and fall short in customized logging activities. 
To overcome such limitation, recent studies employ neural-based methods to decide about \textit{single ingredients} of logging statements (i.e., logging levels, logging variables, logging text). For example, prior work~\cite{li2021deeplv,liu2022tell} predicts the appropriate logging level by feeding surrounding code features to a neural network.
% extracts features from surrounding code and feeds them into a neural network, which can predict an appropriate logging level learning from historically labeled logging statements.
While these tools have also shown improvements in suggesting important variables~\cite{liu2019variables} or proper log levels~\cite{liu2022tell, li2017log}, they lack the ability to produce complete logging statements containing multiple ingredients simultaneously.
Some tools~\cite{li2021deeplv} require the availability of certain ingredients to suggest others, which can be impractical for programmers who need to generate complete logging statements.
However, the complete statement generation has been considered challenging as the model should analyze the code structure, comprehend the developer's intention, and produce meaningful logging text~\cite{mastropaolo2022using}. Moreover, existing neural-based tools are further restricted by training data with limited logging statements and may not generalize to unseen code.

Recent large pre-trained language models (LLMs)~\cite{floridi2020gpt, liu2019roberta} have achieved impressive performance in the field of natural language processing (NLP).
% and offer another solution to generate complete logging statements. 
Inspired by this, the latest logging-specific model, LANCE~\cite{mastropaolo2022using},
treats logging statements generation as a text-to-text generation problem and trains a language model for it. 
% LLMs are motivated by a ``well-read'' human who, after being trained from the vast amount of data, can capture knowledge in its huge model parameters~\cite{han2021pre}. 
LLMs have proven their efficacy in many code intelligence tasks, such as generating functional code~\cite{fried2022incoder, guo2022unixcoder} or resolving bugs~\cite{xia2023automated}, and have even been integrated as plugins for developers~\cite{copilot_research} (e.g., Copilot~\cite{copilot_doc}, CodeWhisperer~\cite{codewhisperer}). 
However, their capacity for generating complete logging statements has not been comprehensively examined.
% implementations as plugins and applications
% Although LLMs have been demonstrated to effectively capture code knowledge, even many of them have been implemented as plugins and applications, and they have become a part of many programmers' daily coding routine, they have never been examined for their potential to generate logging statements. 
To fill this gap, we pose the following question: \textit{To what extent can LLMs produce correct and complete logging statements for developers?}
We expect LLMs, given their strong text generation abilities, can improve the quality of logging statements. Further, LLMs have exhibited a powerful aptitude for code comprehension~\cite{xu2022systematic}, which paves the way for uncovering the semantics of logging variables. 

% Software engineering researchers also employ LLMs as assistant programmers to automatically fix program bugs~\cite{drain2021generating} or write code~\cite{ahmad2021unified}. 
% However, it is restricted by training data with a limited amount of log varieties and may not generalize to unseen code.
% The most recent tool, LANCE~\cite{mastropaolo2022using}, abstracts logging statements generation as a text-to-text generation problem and trains a model to solve it, considering a snippet of code as input text and outputs the code with appropriate logging statements inserted. Nevertheless, either the model size or the training data size is smaller than other LLMs.
% However, LANCE regards code snippets as strings and ignores the code structural information, which has been demonstrated to be helpful in numerous code-related tasks[cite]. 
% The most recent tool, LANCE~\cite{mastropaolo2022using}, abstracts logging statements generation as a text-to-text generation problem and trains an LLM to solve it, considering a snippet of code as input text and outputs the code with appropriate logging statements inserted. However, LANCE regards code snippets as strings and ignores the code structural information, which has been demonstrated to be helpful in numerous code-related tasks[cite]. 

% and write requested functional codes for programmers~\cite{ahmad2021unified}, they have never been investigated for the logging statement generation process. 

\begin{table*}[t]
\centering
\small
\vspace{-0.1in}
\caption{Summarization of key findings and implications in this paper.}
\vspace{-0.1in}
\label{tab:key-summarization}
\begin{tabular}{l||l}
\toprule
   \textbf{Key findings}  &  \textbf{Key implications \& Actionable advice} \\
\toprule   
\makecell[l]{\faHandPointRight[regular]~ The performance of existing LLMs in generating complete log-\\ging statements \textit{needs to be improved} for practical logging usage.} & \makecell[l]{\faArrowAltCircleRight[regular] How to \textit{generate proper logging text} warrants more explo-\\ration.}\\
\makecell[l]{\faHandPointRight[regular]~ Comparing the LLMs' logging capabilities presents a challenge,\\ as models perform inconsistently on different ingredients.} & \makecell[l]{\faArrowAltCircleRight[regular] Intriguing alternative, possibly \textit{unified metrics} to assess the\\ quality of logging statements.}\\ 
\midrule
\makecell[l]{\faHandPointRight[regular]~ Directly applying LLMs yields \textit{better performance} than conv-\\entional logging baselines.} & \multirow{5}{*}{\makecell[l]{\faArrowAltCircleRight[regular] LLM-powered logging is promising. Refining prompts with \\instructions and demonstration selection strategies for effective\\ few-shot learning should be investigated.}}\\ 
\makecell[l]{\faHandPointRight[regular]~ Instructions significantly impact LLMs, but there is consistency\\ in the relative ranking of LLMs when used with same instructions.} \\
\makecell[l]{\faHandPointRight[regular]~ Demonstrations help, but more demonstrations does not always\\ lead to a higher logging performance.} \\
\midrule
\makecell[l]{\faHandPointRight[regular]~ Since comments provide code intentions from developers, ignor-\\ing them leads to decreased effectiveness for LLMs.} & \multirow{3}{*}{\makecell[l]{\faArrowAltCircleRight[regular] Providing proper \textit{programming contexts} over the projects\\ that reveal execution information can boost LLMs' logging\\ performance.}}\\  
\makecell[l]{\faHandPointRight[regular]~ Compared to comments, LLMs gain greater advantages from\\ considering \textit{additional methods} in the same file.} & \\
\midrule
\makecell[l]{\faHandPointRight[regular]~ \textit{Unseen code} significantly degrades all LLMs' performance, par-\\ticularly in variable prediction and logging text generation.} & \makecell[l]{\faArrowAltCircleRight[regular] To advance the generalization capabilities of LLMs, devel-\\oping \textit{prompt-based learning techniques} to capture code logic\\ offers great potential of LLMs in automated logging.} \\
\bottomrule
\end{tabular}
\vspace{-0.1in}
\end{table*}

\textbf{Our work.}
To answer our research question, this empirical study thoroughly investigates how modern LLMs perform logging statement generation from two perspectives: \textit{effectiveness} and \textit{generalization capabilities}. 
% In particular, we evaluate X state-of-the-art LLMs in a newly collected dataset [DATA-NAME]. [DATA] is built on Java and involves 122,990 Java files, [num] methods, and [num] logging statements in total. It is obtained from 14,747 well-maintained projects on GitHub.
We extensively evaluate and understand the effectiveness of LLMs by studying (1) their ability to generate logging ingredients, (2) the impact of input instructions and demonstrations, and (3) the influence of external program information.
% through the following questions:
% \begin{itemize}
%     \item \textbf{RQ1: How do different LLMs perform in deciding ingredients of logging statements generation?}
%     \item \textbf{RQ2: What internal characteristics of LLMs will affect logging generation?}
%     \item \textbf{RQ3: How do external factors influence the effectiveness in generating logging statements?}
% \end{itemize}
To assess the generalizability of LLMs, since LLMs are trained on a significant portion of publicly available code, there is a potential data leakage issue in which logging statements used for evaluation purposes may be included in the original training data~\cite{xia2023automated, rabin2023memorization, jiang2023impact}. It remains unclear whether LLMs are really inferring logging statements or merely memorizing the training data. 
% Promising logging tools must possess strong generalizability to tackle the unseen code.
Thus, we further evaluate the generalization capabilities of LLMs using unseen code. 
% in the following question:
% \begin{itemize}
%     \item \textbf{RQ4: How do different LLMs generate logging statements for unseen code?}
% \end{itemize}
% In this paper, we thoroughly investigate how modern LLMs perform the task of generating logging statements. 

In particular, we evaluate the performance of eleven top-performing LLMs encompassing a variety of types—including natural language and code-oriented models, covering both academic works and commercial coding tools on \textit{\Odata}, a new dataset we collected, consisting of \textit{2,430} Java files, \textit{3,870} methods, and \textit{6,849} logging statements. 
Additionally, we employ a lightweight code transformation technique to generate a semantics-equivalent modified dataset \textit{\Tdata}, which contains previously untrained data and thus can be used to evaluate the generalization capabilities of LLMs. 
Based on our large-scale empirical study on \Odata~and \Tdata, we summarize eight key findings and five implications with actionable advice in Table~\ref{tab:key-summarization}.

\textbf{Contributions.} The contribution of this paper is threefold:
\begin{itemize}
% \begin{itemize}
    \item We build a logging statement generation dataset, LogBench, containing the collection of \textit{6,849} logging statements in \textit{3,870} methods (\Odata), along with their functionally equivalent unseen code after transformation (\Tdata).
    \item We analyze the logging effectiveness of eleven top-performing LLMs by investigating their performance over various logging ingredients, analyzing prompt information that influences their performance, and examining the generalization capabilities of these LLMs with unseen data.
    \item We summarize our results into eight findings and draw five implications to provide valuable insights for future research on automated log statement generation. All datasets, developed tools,  source code, and experiment results are available in a publicly accessible repository\footnote{Available in: \url{
https://github.com/LoggingResearch/LoggingEmpirical}}.
\end{itemize}

    % To our best knowledge, we conduct the first empirical study to explore LLMs' ability on logging statement generation. 
    %%%%% HERE: num = logbench-O + logbench-T
    % We build the first multi-ingredient logging statement generation benchmark dataset (\Odata)~with \textit{3,870} methods and \textit{6,849} logging statements, and their associated unseen dataset~(\Tdata) after code transformation. Using the datasets, we extensively examine the effectiveness and generalization capabilities of LLMs.
    % which can be used to assess the effectiveness and generalization ability of automated logging generators.
    % , including XXX logging statements in 122,990 Java files from 14,747 GitHub projects.
\section{Background}\label{sec:background}

\subsection{Problem Definition}

\begin{figure}[tbp]
     \centering
    \includegraphics[width=0.9\columnwidth]{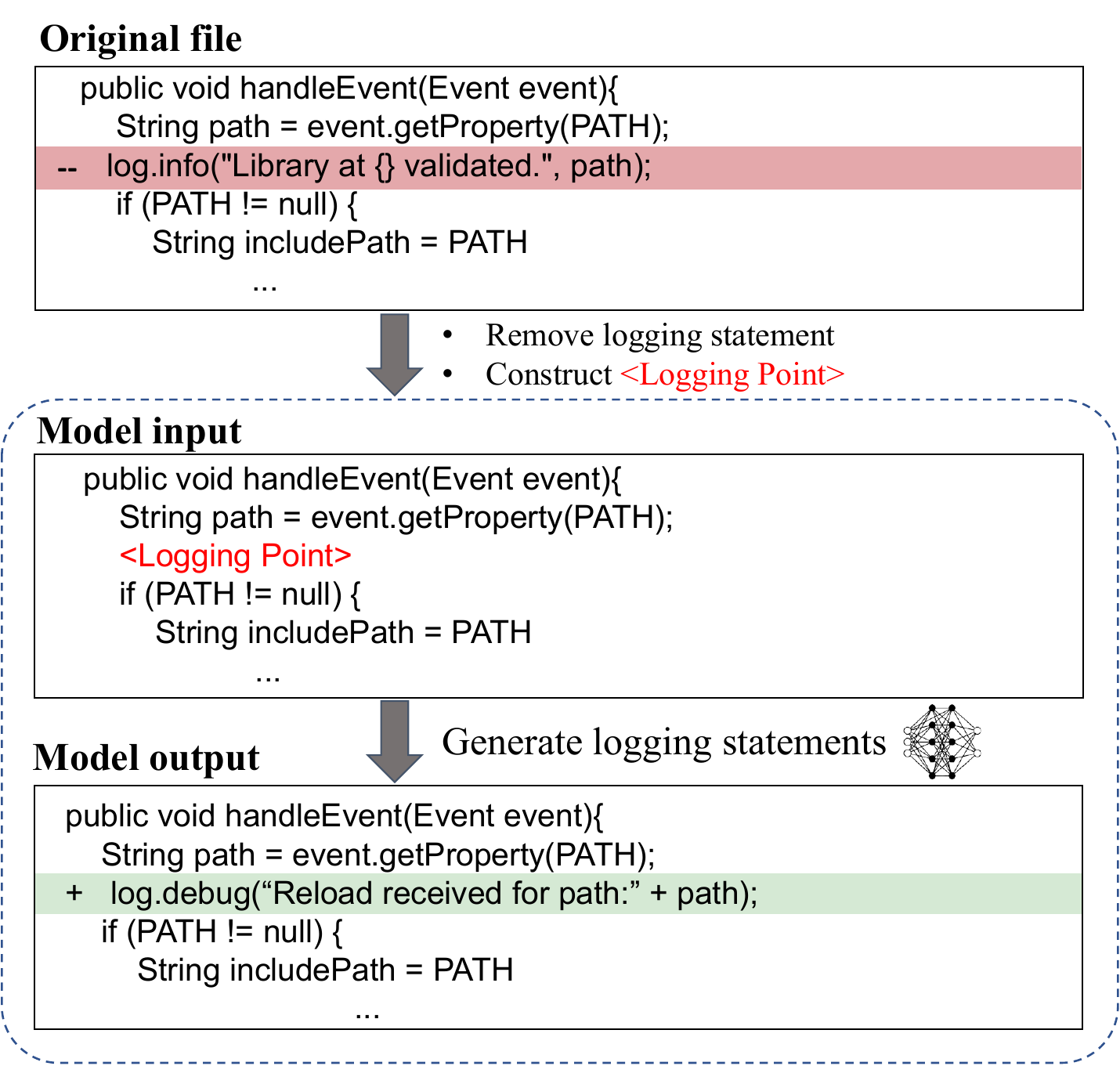}
     \caption{Task formulation: given a method and a specific logging point, the model is asked to predict the logging statement in the point. }
     \label{fig:task-formulation}
 \end{figure}

This study focuses on the \textit{logging statement generation} task (i.e., \textit{what-to-log}), which can be viewed as a statement completion problem: given lines of code (typically a method) and a specific logging point between two statements, the generator is then required to predict the logging statement at such point. 
The prediction is expected to be similar to the one removed from the original file.
Figure~\ref{fig:task-formulation} (in dashed line) illustrates an example of this task, where an effective logging statement generator should suggest \fixedwidth{log.debug("Reload received for path:" + path)} that is highlighted with green for the specified logging point\footnote{In this paper, the logging statement that the generator should predict is always highlighted by \textcolor{green}{green}.}. 
Following a previous study~\cite{mastropaolo2022using}, for the code lines with \textit{n} logging statements, we create \textit{n-1} inputs by removing each of them one at a time.

\begin{table*}[t]
    \centering
    \caption{Study subjects involved in our empirical study.
    % and query reformulation baselines included in this empirical study. For tools we list the year of its most recent update time.
    }
    \begin{tabular}{ccp{8.5cm}ccc}
    \toprule
    \multirow{2}*{\textbf{Model}} & \multirow{2}*{\textbf{Access}} & \multirow{2}*{\textbf{Description}}
    & \multirow{2}*{\tabincell{c}{\textbf{Pre-trained corpus} \\ \textbf{(Data size)}}} & \multirow{2}*{\textbf{\#Params}} & \multirow{2}*{\textbf{Year}} \\
     % \textbf{Model} & \textbf{Access} & \textbf{Description} & \textbf{Pretrained Objectives} & \textbf{\# Params} & \textbf{Year} \\
        % \multirow{2}*{\textbf{Field}} & \multirow{2}*{\tabincell{c}{\textbf{Category/} \\ \textbf{Data Source}}} & \multirow{2}*{\textbf{Description}} &  \multirow{2}*{\textbf{\tabincell{c}{PL}}} & \multirow{2}*{\textbf{Venue}} & \multirow{2}*{\textbf{Year}} \\
     & & & & \\
    \midrule
    \rowcolor{ashgrey}
    \multicolumn{6}{c}{\textbf{General-purpose LLMs}}\\
    \midrule
    \multirow{3}{*}{\tabincell{c}{Davinci}} & \multirow{3}{*}{\tabincell{c}{API}} & Davinci is derived from InstructGPT~\cite{ouyang2022training} is an ``instruct`` model meant to generate texts with clear instructions. We access the Text-davinci-003 model by calling the official API from OpenAI. & \multirow{3}{*}{\tabincell{c}{-}} & \multirow{3}{*}{\tabincell{c}{175B}} & \multirow{3}{*}{\tabincell{c}{2022}} \\
    \midrule
    \multirow{4}{*}{\tabincell{c}{ChatGPT}} & \multirow{4}{*}{\tabincell{c}{API}} & ChatGPT is an enhanced version of GPT-3 models~\cite{gpt-3.5}, with improved conversational abilities achieved through reinforcement learning from human feedback~\cite{christiano2017deep}. It forms the core of the ChatGPT system~\cite{ChatGPT}. We access the GPT3.5-turbo model by calling the official API from OpenAI. & \multirow{4}{*}{\tabincell{c}{-}} & \multirow{4}{*}{\tabincell{c}{175B}} & \multirow{4}{*}{\tabincell{c}{2022}} \\
    \midrule
    \multirow{3}{*}{\tabincell{c}{Llama2}} & \multirow{3}{*}{\tabincell{c}{Model}} & Llama2~\cite{touvron2023llama} is an open-sourced LLM trained on publicly available data and outperforms other open-source conversational models on most benchmarks. We deploy the Llama2-70B model provided by the authors. & \multirow{3}{*}{\tabincell{c}{Publicly available \\ sources \\ (2T tokens)}}  & \multirow{3}{*}{\tabincell{c}{70B}} & \multirow{3}{*}{\tabincell{c}{2023}} \\
    \midrule
    \rowcolor{ashgrey}
    \multicolumn{6}{c}{\textbf{Logging-specific LLMs}}\\
    \midrule
    \multirow{5}{*}{\tabincell{c}{LANCE}} & \multirow{5}{*}{\tabincell{c}{Model}} & LANCE~\cite{mastropaolo2022using} accepts a method that needs one logging statement and outputs a proper logging statement in the right position in the code. It is built on the T5 model, which has been trained to inject proper logging statements. We re-implement it based on the replication package~\cite{LanceReplication} provided by the authors. & \multirow{5}{*}{\tabincell{c}{Selected GitHub \\ projects \\ (6M methods)}}   & \multirow{5}{*}{\tabincell{c}{60M}} & \multirow{5}{*}{\tabincell{c}{2022}} \\
    \midrule
    \rowcolor{ashgrey}
    \multicolumn{6}{c}{\textbf{Code-based LLMs}}\\
    \midrule
    \multirow{4}{*}{\tabincell{c}{InCoder}}  & \multirow{4}{*}{\tabincell{c}{Model}} & InCoder~\cite{fried2022incoder} is a unified generative model trained on vast code benchmarks where code regions have been randomly masked. It thus can infill arbitrary code with bidirectional code context for challenging code-related tasks. We deploy the InCoder-6.7B model provided by the authors. &
    \multirow{4}{*}{\tabincell{c}{GitHub, GitLab,\\ StackOverflow \\ (159GB code, \\ 57GB StackOverflow)}} & \multirow{4}{*}{\tabincell{c}{6.7B}} & \multirow{4}{*}{\tabincell{c}{2022}} \\
    \midrule
    \multirow{3}{*}{\tabincell{c}{CodeGeeX}} & \multirow{3}{*}{\tabincell{c}{IDE Plugin}} & CodeGeeX~\cite{codegeex} is an open-source code generation model, which has been trained on 23 programming languages and fine-tuned for code translation. We access the model via its plugin in VS Code. & \multirow{3}{*}{\tabincell{c}{GitHub code \\ (158.7B tokens)}}  & \multirow{3}{*}{\tabincell{c}{13B}} & \multirow{3}{*}{\tabincell{c}{2022}} \\
    % selected from the Pile, the CodeParrot dataset, and additional data for Python, Java, and C++.  
    \midrule
    \multirow{4}{*}{\tabincell{c}{StarCoder}}  & \multirow{4}{*}{\tabincell{c}{Model}} & StarCoder~\cite{li2023starcoder} has been trained on 1 trillion tokens from 80+ programming languages, and fine-tuned on another 35B Python tokens. It outperforms every open LLM for code at the time of release. We deploy the StarCoder-15.5B model provided by the authors. & \multirow{4}{*}{\tabincell{c}{The Stack \\ (1T tokens)}}  & \multirow{4}{*}{\tabincell{c}{15.5B}} & \multirow{4}{*}{\tabincell{c}{2023}} \\
    \midrule
    \multirow{4}{*}{\tabincell{c}{CodeLlama}} & \multirow{4}{*}{\tabincell{c}{Model}} & CodeLlama~\cite{roziere2023code} is a family of LLMs for code generation and infilling derived from Llama2. After they have been pretrained on 500B code tokens, they are all fine-tuned to handle long contexts. We deploy the CodeLlama-34B model provided by the authors. & \multirow{4}{*}{\tabincell{c}{Publicly available \\ code \\ (500B tokens)}}  & \multirow{4}{*}{\tabincell{c}{34B}} & \multirow{4}{*}{\tabincell{c}{2023}}\\
    \midrule
    \multirow{4}{*}{\tabincell{c}{TabNine}} & \multirow{4}{*}{\tabincell{c}{IDE Plugin}} & TabNine~\cite{tabnine} is an AI code assistant that can suggest the following lines of code. It can automatically complete code lines, generate entire functions, and produce code snippets from natural languages. We access the model via its plugin in VS Code. & \multirow{4}{*}{\tabincell{c}{-}}& \multirow{4}{*}{\tabincell{c}{-}}& \multirow{4}{*}{\tabincell{c}{2022}}\\
    \midrule
    \multirow{4}{*}{\tabincell{c}{Copilot}} & \multirow{4}{*}{\tabincell{c}{IDE Plugin}} & Copilot~\cite{copilot_research} is a widely-studied AI-powered code generation tool relying on the CodeX~\cite{codex}. It can extend existing code by generating subsequent code trunks based on natural language descriptions. We access the model via its plugin in VS Code.& \multirow{4}{*}{\tabincell{c}{-}}& \multirow{4}{*}{\tabincell{c}{-}}& \multirow{4}{*}{\tabincell{c}{2021}} \\
    \midrule
    \multirow{4}{*}{\tabincell{c}{CodeWhisperer}} & \multirow{4}{*}{\tabincell{c}{IDE Plugin}} & CodeWhisperer~\cite{codewhisperer}, developed by Amazon, serves as a coding companion for software developers. It can generate code snippets or full functions in real-time based on comments written by developers. We access the model via its plugin in VS Code.& \multirow{4}{*}{\tabincell{c}{-}}& \multirow{4}{*}{\tabincell{c}{-}}& \multirow{4}{*}{\tabincell{c}{2022}}\\
    \bottomrule

    \label{tab:llms-summary}
    \end{tabular}
\end{table*}

% \begin{table}[t!]
% \small
%     \centering
%     \caption{Model summarization}
%     \vspace{-0.1in}
%     \label{tab:model-summary}
%     \resizebox{\linewidth}{!}{%
%     \begin{NiceTabular}{lcccc}
%     \CodeBefore
%             \rowcolors{2}{}{gray!15}
%             \Body
%         \toprule
%            \textbf{Model} & \textbf{Access} & \textbf{Task} & 
%             \textbf{\# Params} &
%             \textbf{Year}\\
%         \midrule
%                     Davinci & API & Instruction following & 175B & 2022
%             \\
%             ChatGPT& API & AI chatbot
%              & 175B
%              & 2022
%              \\
%             LANCE & Model & Logging statement recommendation
%              & 60M
%              & 2022
%              \\
%              InCoder& Model & Code infilling and synthesis & 6.7B& 2022 \\
%             CodeGeex & Plugin & Code completion
%              & 13B
%              & 2022
%              \\
%             TabNine& Plugin & Code completion & - & 2022 \\
%             Copilot$^\dagger$ & Plugin & Code completion
%              & -
%              & 2021
%              \\
%             CodeWhisperer & Plugin & Code completion & - &2022 \\
%         \bottomrule    
%         \multicolumn{4}{l}{\footnotesize $^\dagger$ Copilot is powered by Codex with 12B parameters~\cite{DBLP:journals/corr/abs-2111-03922}.} \\
%         \end{NiceTabular}}
%         % \multicolumn{5}{l}{\parbox{\linewidth}{$^\dagger$ Copilot is powered by Codex with 12B parameters~\cite{DBLP:journals/corr/abs-2111-03922}.}}
%     %
%                 \vspace{-0.2in}
% \end{table}
\subsection{Challenges in Logging Statement Generation}
\begin{table*}[t]
    \centering
    \caption{Conventional logging approach for single ingredient recommendations.
    }
    \label{tab:conventional-summary}

    \begin{tabular}{ccp{9.5cm}ccc}
    \toprule
    \multirow{2}*{\textbf{Ingredient}} & \multirow{2}*{\textbf{Model}} & \multirow{2}*{\textbf{Description}} & \multirow{2}*{\textbf{\#Params}} & \multirow{2}*{\textbf{Venue}} & \multirow{2}*{\textbf{Year}} \\
    & & & \\
    \midrule
    \multirow{3}{*}{\tabincell{c}{\cellcolor{ashgrey}Logging\\ \cellcolor{ashgrey}levels}} & \multirow{3}{*}{\tabincell{c}{DeepLV}} & DeepLV~\cite{li2021deeplv} leverages syntactic context and message features of the logging statements extracted from the source code to make suggestions on choosing log levels by feeding all the information into a deep learning model. We reimplement the model based on the replication package provided by the authors*. & \multirow{3}{*}{\tabincell{c}{0.2M}} & \multirow{3}{*}{\tabincell{c}{ICSE}} & \multirow{3}{*}{\tabincell{c}{2021}} \\
    \midrule
    \multirow{4}{*}{\tabincell{c}{\cellcolor{ashgrey}Logging\\\cellcolor{ashgrey}Variables}} & \multirow{4}{*}{\tabincell{c}{WhichVar}} & WhichVar~\cite{liu2019variables} applies an RNN-based neural network with a self-attention mechanism to learn the representation of program tokens, then predicts whether each token should be logged through a binary classifier. We reimplement the model based on its paper due to missing code artifacts*. & \multirow{4}{*}{\tabincell{c}{40M$^\dagger$}} & \multirow{4}{*}{\tabincell{c}{TSE}} & \multirow{4}{*}{\tabincell{c}{2021}} \\
    \midrule
     \multirow{5}{*}{\tabincell{c}{\cellcolor{ashgrey} Logging \\ \cellcolor{ashgrey}Text}} & \multirow{5}{*}{\tabincell{c}{LoGenText-Plus}} & LoGenText-Plus~\cite{ding2023logentextplus} generates the logging texts by neural machine translation models (NMT). 
     It first extracts a syntactic template of the target logging text by code analysis, then feeds such templates and source code into Transformer-based NMT models. We reproduce the model based on the replication package provided by the authors. & \multirow{5}{*}{\tabincell{c}{22M}}  & \multirow{5}{*}{\tabincell{c}{TOSEM}} & \multirow{5}{*}{\tabincell{c}{2023}} \\
    \bottomrule
    \end{tabular}
    \begin{flushleft}
    $^\dagger$ The number of parameters (40M) includes the embedding module of the model.
    
    * All the baselines we have reimplemented has been organized in our artifacts..
    \end{flushleft}
\end{table*}

% challenging in understanding code function, software status
The composition of logging statements naturally makes the logging generation problem a joint task of code comprehension and text generation. 
Compared to code completion tasks, the generation of logging statements presents two distinct challenges: (1) inference of critical software  runtime status and (2) the creation of complicated text that seamlessly integrates both natural language and code elements.

First, while code generation produces short methods with a high degree of functional similarity, logging statements are \textit{non-functional} statements not discussed in code generation datasets (e.g., HumanEval~\cite{chen2021codex}, APPS~\cite{hendrycks2021measuring}).
Nevertheless, logging statements are indispensable in large-scale software repositories for documenting run-time system status.
% Logging every status increases performance overhead, whereas logging insufficient status obstacles system debugging. 
To log proper system status, a logging statement generator shall comprehend program structure (e.g., exception handling) and recognize critical code activities worthy of logging.
Second, integrating natural language text and code variables poses a unique challenge. Logging statement generators must be mastered in two distinct languages and harmoniously aligned. Developers describe code functionalities in natural language and then incorporate relevant logging variables.
Likewise, a logging statement generator should be capable of translating runtime code activities into natural language and explaining and recording specific variables.

% Logging statement generation problem requires different (1) code understanding capabilities and (2) code generation abilities than classical code completion tasks, due to their different working subjects.
% The logging statement generation problem has a different subject than the code generation problem, as the latter concentrates on producing the method with high code functional similarity. Existing code generation dataset (Codexglue, )
% challenges: (1) understanding the vital source code activities, (2) generating proper descriptive texts containing both natural language and code elements.

% challenging in understanding developers' opinions.

\subsection{Study Subject}

Motivated by the code-related text generation nature of the logging statement generation, we opt to investigate top-performing LLMs from three fields as our study subjects: LLMs designed for general natural text generation, LLMs tailored for logging activities, and LLMs for code intelligence. We also evaluate state-of-the-art logging suggestion models, which usually work on a single ingredient, to discuss whether advanced LLMs outperform conventional ones. 

We summarize the details of eleven LLMs in Table~\ref{tab:llms-summary} and three conventional approaches in Table~\ref{tab:conventional-summary}. Since we already included official models~\cite{codex,ChatGPT,gpt-3.5} from the GPT series, other models that have been tuned on GPT~\cite{black2022gpt,gpt-j} are not included in our study (e.g., GPT-Neo~\cite{black2022gpt} and GPT-J~\cite{gpt-j}). 
% In short, we evaluate the logging effectiveness of eleven LLMs varying from \yt{} to \yt{} parameters and three 
% We evaluate the effectiveness of state-of-the-art LLMs varying from 60M to 175B parameters.
% % , which are shown to be superior in natural language understanding and code comprehension.
% We summarize their access categories (Access), proposed tasks (Task), and number of parameters (\# Params) in Table~\ref{tab:model-summary}. 
% % These LLMs can be accessed and utilized in various ways, which we summarized them in Table~\ref{tab:model-summary}. 
% % , including through IDE plugins and by downloading the released models
% Since we already included official models~\cite{codex,ChatGPT,gpt-3.5} from the GPT series, other models that have been tuned on GPT~\cite{black2022gpt,gpt-j} are not included in our study (e.g., GPT-Neo~\cite{black2022gpt} and GPT-J~\cite{gpt-j}). 
% % In addition, other logging models~\cite{liu2022tell,li2021deeplv,he2018characterizing,dai2022reval,liu2019variables} predict exclusively single ingredients by utilizing information from other ingredients.
% % These models could introduce bias and are not suitable for this study.
% The investigated LLMs are as follows.

\subsubsection{General-purpose LLMs}
The GPT-series models are designed to produce natural language text closely resembling human language.  
% These models boast large amounts of parameters and undergo a two-step process: pre-training on vast volumes of textual data followed by fine-tuning for specific downstream tasks.
The recent GPT models have demonstrated exceptional performance, dominating numerous natural language generation tasks, such as question-answering~\cite{tan2023can} and text summarization~\cite{goyal2022news}.
Recently, Meta researchers built an open model, LLaMa, as a family member of LLMs~\cite{touvron2023llama}, which showed more efficient and competitive results with GPT-series models.
In our paper, we select the two most capable GPT-series models based on previous work~\cite{ye2023comprehensive}, i.e., Davinci, ChatGPT for evaluation. We also select one competitive open-sourced model, Llama2, as the representative of general-purpose LLMs.
% \begin{itemize}
%     \item \textit{Text-davinci-003 (denoted as Davinci)} is derived from InstructGPT~\cite{ouyang2022training}, which is an ``instruct'' model that is meant to generate texts with clear instructions. 
%     \item \textit{GPT3.5-turbo (denoted as ChatGPT)} is an enhanced version of GPT-3 models~\cite{gpt-3.5}, with improved conversational abilities achieved through reinforcement learning from human feedback~\cite{christiano2017deep}. It forms the core of the ChatGPT system~\cite{ChatGPT} and is among the most advanced language models to date.
%     \item \textit{LLaMa2}~\cite{touvron2023llama} is an open-sourced LLM trained on publicly available data and outperforms other open-source conversational models on most benchmarks.
% \end{itemize}

\subsubsection{Logging-specific LLMs}
To the best of our knowledge, LANCE~\cite{mastropaolo2022using} is the only work on training LLMs for automatically generating logging statements, which has been published in top-tier software venues (i.e., FSE, ICSE, ASE, ISSTA, TSE, and TOSEM). Consequently, we choose it as logging-specific LLMs.
% \begin{itemize}
%     \item \textit{LANCE} is the state-of-the-art model for generating complete logging statements in code. It accepts a method that needs one logging statement and outputs a meaningful log message with a proper logging level in the right position in the code. It is built on the Text-To-Text-Transfer-Transformer model, which has been trained with the goal of injecting proper logging statements.
% \end{itemize}

\subsubsection{Code-based LLMs}
Inspired by the considerable success of LLMs in the natural language domain, researchers also derive Code-based LLMs that can support code understanding and generation tasks, so as to assist developers in completing codes. These LLMs are either commercial models powered by companies, or open-access models in academia. 
For the open-access models with publicly available weights, we follow the selection of code models on recent comprehensive evaluation studies~\cite{roziere2023code, li2023starcoder, zan2023large}, and reserve the LLMs with larger sizes than 6B. The process leads to four LLMs as our subjects, i.e., InCoder~\cite{fried2022incoder}, CodeGeex~\cite{codegeex}, StarCoder~\cite{li2023starcoder}, and CodeLlama~\cite{roziere2023code}.
In terms of the commercial models, we select three popular developer tools as the study subjects, i.e., TabNine~\cite{tabnine}, Copilot~\cite{copilot_research}, and CodeWhisperer~\cite{codewhisperer} from Amazon.

\subsubsection{Conventional Logging Approaches}
Apart from LLMs that can offer complete logging statements, we also select conventional logging approaches that work on \textit{single} logging ingredients for comparison. Specifically, for each ingredient, 
we choose the corresponding state-of-the-art logging approaches from the top-tier software venues: DeepLV~\cite{li2021deeplv} for log level prediction, Liu et al.'s~\cite{liu2019variables} (denoted as WhichVar) for logging variable prediction, and LoGenText-Plus~\cite{ding2023logentextplus} for logging text generation. These approaches learn the relationships between specific logging ingredients and the corresponding code features based on deep learning techniques. Details are summarized in Table~\ref{tab:conventional-summary}.
% When developers wish to log system activities, they must identify the logging point before determining the content of the logging statement. Therefore, it is logical to 
% ask the logging statement generator to complete the logging statement given a specified logging point. 

% employ an automated logging statement generation model to 

% \yt{consider other terminologies in ``background subsection''. Also, use notations.}
% Although previous approaches have been proposed to solve one sub-task of the logging statement automation (e.g., deciding log-or-not, suggesting the log level), these sub-tasks are not sufficient to simulate what programmers face when they write logs. To explore how AI programmers assist in automated writing logs, we formulate the logging statement automation task as follows.

% For logging statement automation, the model considers a code method as input and outputs the complete logging statement in the method. Considering a developer who wants to log system activities, he/she will first determine whether a log point is needed, then decide the logging statement content. Hence, it is natural to ask the logging statement automation model to complete the task with two phases sequentially: where-to-log and what-to-log.

% In particular, <Use a figure>

% In our study, we designed three settings for logging statement generation.

% 1) Logging statement generation/infilling: the logging location is known and the goal is to generate the logging statement given the prefix and suffix of the context.

% 2) Logging statement insertion

\begin{figure*}[tbp]
    \centering
    \includegraphics[width=\linewidth]{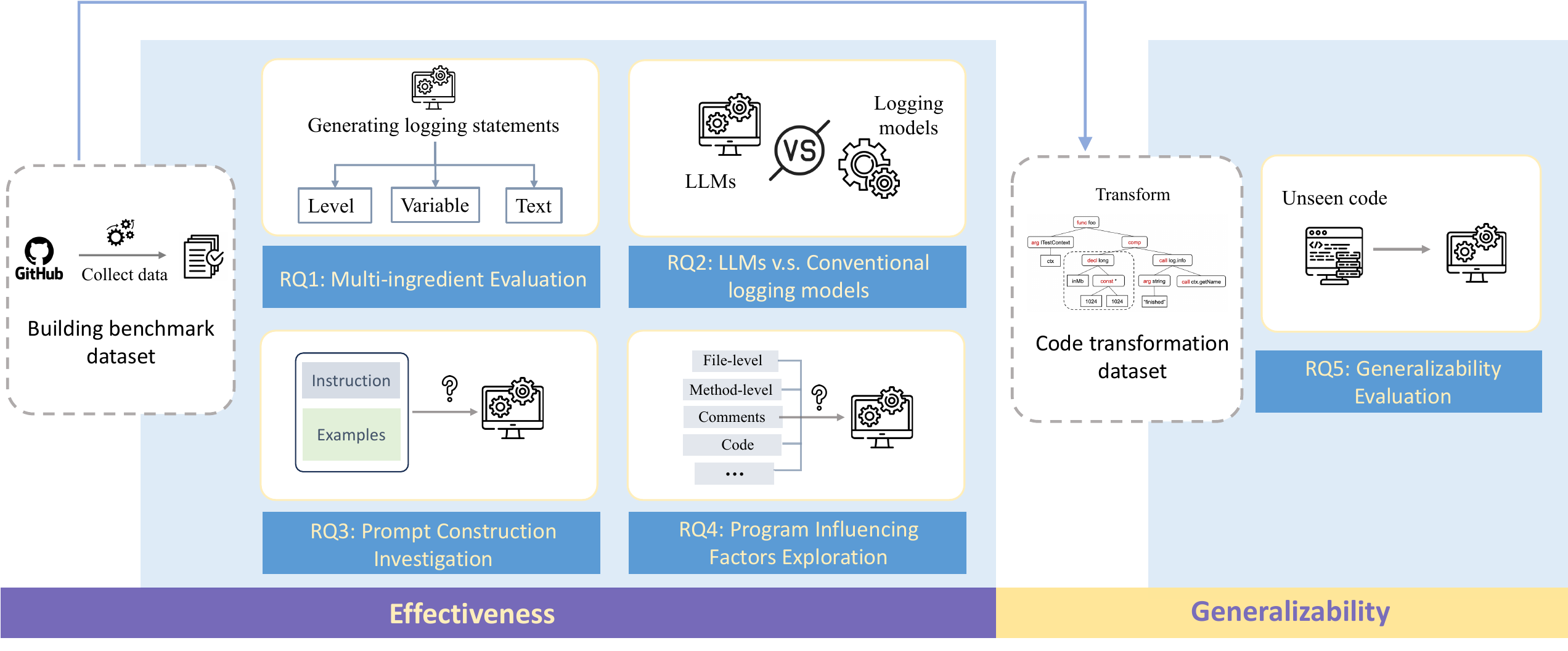}
    \vspace{-0.2in}
    \caption{The overall framework of this study involving five research questions.}
    \label{fig:framework}
\end{figure*}

\section{Study Methodology}\label{sec:methodology}

\subsection{Overview}

Fig.~\ref{fig:framework} depicts the overview framework of this study involving five research questions from two perspectives: (1) \textit{effectiveness}: how do LLMs perform in logging practice?
and (2) \textit{generalizability}: how well do LLMs generate logging statements for unseen code?

To start, we develop a benchmark dataset \Odata~comprising \textit{6,849} logging statements in \textit{3,870} methods by crawling high-quality GitHub repositories. Inspired by the success of LLMs in NLP and code intelligence tasks, our focus is on assessing their efficacy in helping developers with logging tasks.
This study first evaluates the effectiveness of state-of-the-art LLMs in terms of multiple logging ingredients (RQ1).
We then conduct a comparative analysis between state-of-the-art conventional logging tools and LLMs, elucidating differences and providing insights into potential future model directions (RQ2).
Next, we investigate the impact of instructions and demonstrations as inputs for LLMs, offering guidance for effectively prompting LLMs for logging (RQ3).
Furthermore, we investigate how external influencing factors can enhance LLM performance, identifying effective program information that should be input into LLMs to improve logging outcomes (RQ4).
Last but not least, we explore the generalizability of LLMs to assess their behavior in developing new and unseen software.
To this end, we evaluate models on an unseen code dataset, \Tdata, which contains code derived from \Odata~ that was transformed to preserve readability and semantics (RQ5).

\subsection{Benchmark Datasets}\label{sec:datasets}
Due to the lack of an existing dataset that can meets the benchmark requirements, we developed the benchmark dataset
\Odata~and \Tdata for logging statement generation in this section. Although we chose Java as the target language of our study, due to its wide presence in industry and research~\cite{chen2020studying}, the experiments and findings can be extended to other programming languages. 

% \ry{how about using a table to show some statistics of our two datasets? Now that information is too dispersive.}

\begin{table*}[tbp]
\centering
\small
\caption{Our code transformation tools with eight code transformers, descriptions, and associated examples.}
\label{tab:codeTran}
  \begin{tabular}{cll}
            \toprule
            Transformer &  Descriptions &  Example\\
            \midrule
            Condition-Dup & Add logically neutral elements (e.g., \textit{\&\& True} or \textit{\textbar \textbar ~False}) & \colorbox{githubred}{if (exp0)} $\to$ \colorbox{githubgreen}{if (exp0 $||$ false)}\\
            Condition-Swap & Swap the symmetrical elements of condition statements & \colorbox{githubred}{if (var0 != null)} $\to$ \colorbox{githubgreen}{if (null != var0)}\\
            Local variable & Extract constant values and assign them to local variables & \colorbox{githubred}{var0 = const0;} $\to$ \colorbox{githubgreen}{int var1 = const0; var0 = var1;}\\
            Assignment &  Separate variable declaration and assignment & \colorbox{githubred}{int var0 = var1;} $\to$ \colorbox{githubgreen}{int var0; var0 = var1;}\\ 
            Constant & Replace constant values with equivalent expressions & \colorbox{githubred}{int var0 = const0} $\to$ \colorbox{githubgreen}{int var0 = const0 + 0}\\For-While & Convert \textit{for-loops} to equivalent \textit{while-loops} & \colorbox{githubred}{for (var0 = 0; var0 $<$ var1; var0++) \{\}} $\leftrightarrow$ \\
             While-For & Convert \textit{while-loops} to equivalent \textit{for-loops} & \colorbox{githubgreen}{var0 = 0; while (var0++ $<$ var1) \{\}} \\
            Parenthesis & Add redundant parentheses to expression & \colorbox{githubred}{var0 = arithExpr0} $\to$ \colorbox{githubgreen}{var0 = (arithExpr0)}\\
            \bottomrule
\end{tabular}
\end{table*}

\subsubsection{Creation of \Odata}
We build a benchmark dataset, consisting of high-quality and well-maintained Java files with logging statements,
by mining open-source repositories from GitHub. As the largest host of source code in the world, GitHub contains a great number of repositories that reflect typical software development processes. 
In particular, we begin by downloading high-quality Java repositories that meet the following requirements\footnote{All repositories were archived on July 2023}:
\begin{itemize}
    \item Gaining more than \textit{20} stars, which indicates a higher level of attention and interest in the project.
    \item Receiving more than \textit{100} commits, which suggests the project is actively maintained and not likely to be disposable.
    \item Engaging with at least \textit{5} contributors, which demonstrates the quality of its logging statements by simulating the collaborative software development environment.
\end{itemize}

We then extract the files that contain logging statements in two steps. We first select the projects whose POM file includes popular logging utility dependencies (e.g., Log4j, SLF4J), resulting in \textit{3,089} repositories.  We then extract the Java files containing at least one logging statement by matching them with regular expressions~\cite{chen2018automated}, because logging statements are always written in specified syntax (e.g., \fixedwidth{log.info()}).
% This step eventually leads to 122,900 Java files and \yt{num} methods in total.
% As commercial coding assistant tools (such as Copilot and TabNine) have restricted usage patterns, we are unable to evaluate them using automated scripts~\cite{terms}.
Afterward, we randomly sample the collected files across various repositories, resulting in a dataset of \textit{2,420} files containing \textit{3,870} methods and \textit{6,849} logging statements, which we refer to as \Odata.
% can only manually use them for this study as previous work~\cite{pearce2022asleep, mastropaolo2023robustness}.
% Therefore, we conduct random sampling on the extracted files across different repositories to enable manual evaluation, resulting in 2420 files consisting of 3870 methods, denoted as \Odata.

% we examine the POM file of each acquired repository to ensure the dependency on log utilities and remove any files that do not contain logging statements. 
% We accomplish this by reserving only those projects whose POM file includes popular logging utilities (e.g., log4j, slf4j), which results in 3,089 repositories.  
% Afterwards, we extract the Java files containing at least one logging statement by matching them with regular expressions as previous work~\cite{chen2018automated}, because logging statements are always written in specified syntax (e.g., \fixedwidth{log.info()}, \fixedwidth{log.error()}). 
% This step leads to 122,900 Java files, \yt{xx} methods, and \yt{xx} code lines in total.

% Afterward, considering logging statements are always written in specified syntax (e.g., \fixedwidth{log.info()}, \fixedwidth{log.error()}), we utilize regular expressions to match the logging statement within each Java file in the selected repositories as previous work~\cite{chen2018automated}. 
% The file with at least one matched logging statement will be reserved as a candidate for \Odata. 
% The Java file with at least one matched logging statement will be reserved for \Odata, leading to 122,900 Java files, \yt{xx} methods, and \yt{xx} code lines in total.

\subsubsection{Creation of \Tdata~Dataset to Avoid Data Leakage}\label{sec:Tdata-creation}

LLMs deliver great performance in multiple tasks; however, evaluating their performance solely on publicly available data can be problematic. Since LLMs are trained on datasets that are obtained through large-scale web scraping~\cite{gao2020pile}, these models may have already seen the benchmark data during their training, raising concerns about assessing their generalization abilities~\cite{xia2023automated, rabin2023memorization, jiang2023impact}.
This issue, commonly known as \textit{data leakage}, requires particular attention since most code models~\cite{fried2022incoder} have been trained on public code. 
% Besides, other models we analyzed are also likely to have seen the public repositories that can be crawled from the Internet.
% \yc{TBD: The training datasets of other models will inevitably also make use of public repositories.}

 \begin{figure}[tbp]
     \includegraphics[width=\columnwidth]{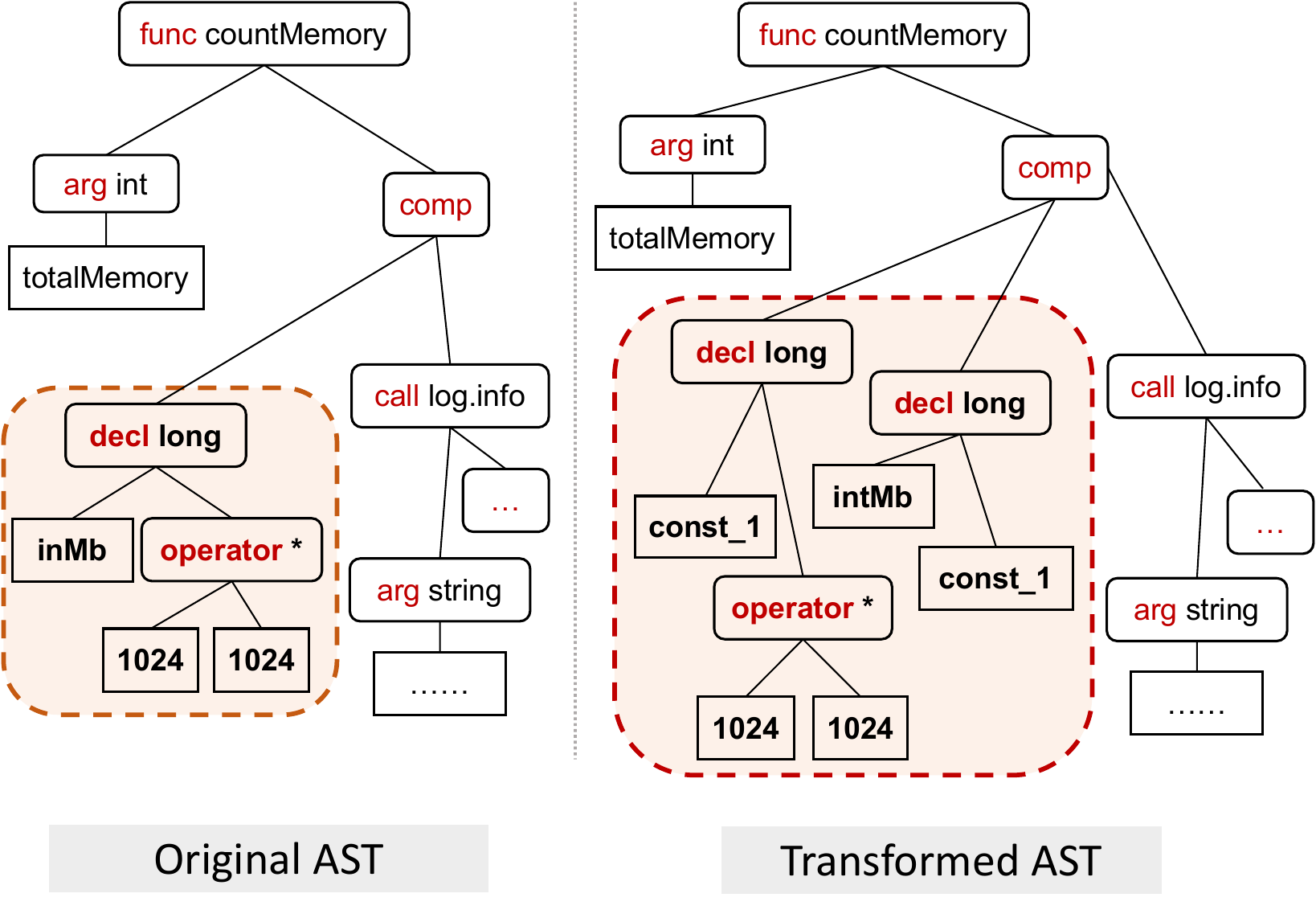}
     \vspace{-0.1in}
     \caption{An example of how the code (constant) transformer works. The constant checker firstly detects transformation points, then the Local variable transformer replaces the constant expression \fixedwidth{\{inMb=1024*1024\}} by \fixedwidth{\{const\_1=1024*1024; inMb=const\_1\}} involving a new variable \textit{const\_1}. The AST changes via transformation are highlighted in \textcolor{red}{red} area.}
     \label{fig:ast-transform}
 \end{figure}

% Furthermore, other code-related datasets, such as BigQuery~\yc{cite} and Pile~\yc{cite}, inevitably contain public code, further emphasizing the need to scrutinize data leakage when evaluating the performance of these models.

% there is concern about their memorization and generalization ability. 

% , because the public evaluation dataset might be included in their training code files~\cite{xia2023automated, rabin2023memorization, jiang2023impact}. 
% As the 
% This \textit{data leakage} issue requires particular attention when LLMs are trained on the datasets that come from the large-scale scraping of the Internet~\cite{gao2020pile}, as the LLMs might have been exposed to the evaluation code files crawled from GitHub during their training phase.

% This issue is especially relevant when training LLMs on large datasets like BigQuery~\yc{cite} or Pile~\yc{cite}, as they might have already been exposed to evaluation code files crawled from public repositories, leading to \textit{data leakage} problems. \yc{changed}

To fairly evaluate the generalization ability of LLMs, we further develop an unseen code dataset \Tdata~that consists of the code transformed from \Odata. 
Prior works have developed \textit{semantics-preserving} code transformation techniques that do not change the functionality of the original code, for the purpose of evaluating the robustness of code models~\cite{quiring2019misleading, li2022closer, li2022cctest,li2021towards}. 
However, these approaches randomly replace informative identifiers with meaningless ones, degrading the \textit{readability} of the code. 
For example, after transforming an informative variable name (e.g., \fixedwidth{totalMemory}) to a non-informative name (e.g., \fixedwidth{var0}), even a programmer can hardly understand the variables and log properly.
Such transformations make the transformed code less likely to appear in daily programming and not suitable for logging practice studies.
To avoid this issue, we devise a code transformation tool that generates semantics-preserving and readability-preserving variations of the original code.

In particular, our code transformation tool employs eight carefully engineered, lightweight code transformers motivated by previous studies~\cite{quiring2019misleading, li2022cctest, donaldson2017automated,cheers2019spplagiarise}, whose descriptions, together with their examples, are illustrated in Table~\ref{tab:codeTran}. These code transformation rules work at the Abstract Syntax Tree (AST) level, ensuring that the transformed code remains semantically equivalent to the original code. Besides, readability-degrading transformations, such as injecting dead code~\cite{balakrishnan2005code} and modifying the identifier names, are eliminated. 
Additionally, to affirm the soundness of our transformations, we have limited our selection to widely-used transformation rules that have been proven effective in various code-related tasks~\cite{li2021towards, quiring2019misleading, zhang2023statfier} over time. Transformation rules are further verified by executing unit tests on sample projects, which confirm that our code transformations will not hurt functionality.
% In particular, our code transformation tool employs a series of carefully engineered, lightweight code transformation rules at the Abstract Syntax Tree (AST) level, ensuring that the transformed code remains semantically equivalent to the original code. 
% Following previous studies~\cite{quiring2019misleading, li2022cctest, donaldson2017automated,cheers2019spplagiarise}, we propose eight different types of code transformers, whose descriptions and examples are shown in Table~\ref{tab:codeTran}. %including but not limited to the variable transformer and statement transformer. 
% Besides, the readability-degrading transformations, such as injecting dead code~\cite{balakrishnan2005code} and modifying the identifier name, are eliminated.
% identifier name modification and dead code injection, are eliminated.

%The procedure of transformation goes as follows: take the source code $x$ and the target transformed version $x'$. 
The process of transformation begins with converting the source code into an AST representation using JavaParser~\cite{javaparser}. To detect potential transformation points (i.e., specific nodes and subtrees) for each transformer, 
a series of predefined checkers traverse the AST in a top-down manner.
Once the transformation points are identified, each checker will independently call its corresponding transformer to perform a one-time transformation. We denote one-time transformation as $T: x \rightarrow x'$, where $x$ and $x'$ represent the source AST and the transformed AST, respectively.
Each transformer functions independently, allowing multiple transformations to be applied to the same code snippet without conflicts. 
These single transformations are chained together to form the overall transformation: $\mathbb T = T_1 \circ T_2 \circ ... \circ T_n $.
Once all the identified points have been transformed or the number of transformations reaches a predetermined threshold, the AST is converted back into the source code to complete the transformation process. 
Fig.~\ref{fig:ast-transform} exhibits a case concerning how a \fixedwidth{Local variable} transformer works. 

\subsection{Implementations}
\subsubsection{Evaluation}
Based on the access ways of different LLMs (Table~\ref{tab:llms-summary}), we evaluated them as follows.
%implemented
% As shown in Table~\ref{tab:model-summary}, LLMs can be accessed in different ways, i.e., shareable model weights, APIs, and plugins of IDEs. We implement the

(1) Released models (Llama2, LANCE, InCoder, StarCoder, CodeLlama): we ran them on a \textit{32}-Core workstation with an Intel Xeon Platinum \textit{8280} CPU, \textit{256} GB RAM, and \textit{4}x NVIDIA GeForce RTX \textit{4090} GPUs in Ubuntu \textit{20.04.4} LTS, using the default bit precision settings for each model.

(2) APIs (ChatGPT, Davinci): we called their official APIs to generate the logging statement by providing the following instruction: 
\textit{Please complete the incomplete logging statement at the logging point: [Code with corresponding logging point]}. As we discussed in Sec. \ref{sec:rq3}, we choose the median value of all metrics across the top five instructions, as determined by voting, to approximate the instructions most commonly utilized by developers. We set its temperature to 0 so that ChatGPT would generate the same output for the same query to ensure reproducibility. For ChatGPT and Davinci, we use the public APIs provided by OpenAI with \textit{gpt-3.5-turbo-0301} and \textit{text-davinci-003}, respectively. 

(3) Plugins (Copilot, CodeGeeX, TabNine, CodeWhisperer): we purchased accounts for each author to obtain the logging statement manually at the logging point that starts with the original logging API (e.g., \fixedwidth{log.}). This starting point forces these plugins to generate logging statements instead of other functional codes. 

For conventional logging approaches, we reproduced them based on the replication packages released by the authors, or the paper descriptions if the replication package is missing.
For all experiments that may introduce randomness, to avoid potential random bias, we repeat them three times and report the median results following previous works~\cite{khan2022guidelines,xu2023prompting,huang2023not}.

\subsubsection{Code Transformation}
Our code transformation technique (Sec. \ref{sec:Tdata-creation}) was implemented using \textit{4,074} lines of Java code, coupled with the JavaParser library~\cite{javaparser}, a widely-used parser for analyzing, transforming, and generating Java code.
All transformations were performed on the same workstation as in the evaluation.

\section{Result analysis}\label{sec:exp}

\subsection{Metrics}

In line with prior work~\cite{he2021survey}, we evaluate the logging statement generation performance concerning three ingredients: \textit{logging levels}, \textit{logging variables}, and \textit{logging texts}.
Although different ingredients emphasize various aspects of runtime information, they are indispensable and complementary resources for engineers to reason about system behavior.
% we identify three key ingredients in effective logging statements based on their composition:
% (1) \textit{Logging levels}~\cite{li2021deeplv,liu2022tell}: indicates the severity of a log event, helping developers to filter logs during system maintenance;
% (2) \textit{Logging variables}~\cite{liu2019variables,lidid,li2017log}: contains essential run-time information from system states; and
% (3) \textit{Logging texts}~\cite{mastropaolo2022using,ding2022logentext,lal2016logoptplus}: provides a description of the system activities and has been most frequently used for system maintenance. 
% Taking \fixedwidth{log.warn("Failed to connect to host: " + URL)} as an example, \fixedwidth{warn} and \fixedwidth{URL} refer to logging level and logging variables, respectively, where logging text is \fixedwidth{Failed to connect to host: <*>}.
% Although different ingredients offer diverse information for system maintainers, they serve as indispensable resources for reasoning the software behavior. 

\begin{table}[tbp]
    \centering
    \vspace{-0.1in}
    \caption{The effectiveness of LLMs in predicting logging levels and logging variables.}
        \vspace{-0.1in}
    \label{tab:rq1-result-level-variable}
    \resizebox{\linewidth}{!}{%
        \begin{NiceTabular}{l||cc|ccc}
        \CodeBefore
            % \rowcolors{3}{gray!15}{}
        \Body
            \toprule
                \Block{3-1}{\textbf{Model}} &
                \Block[c]{1-2}{\textbf{Logging Levels}} & &
                \Block[c]{1-3}{\textbf{Logging Variables}} &
                &
                \\
            \cmidrule{2-6}
                &
                \textbf{L-ACC} &
                \textbf{AOD} &
                \textbf{Precision} &
                \textbf{Recall} &
                \textbf{F1} \\
            \midrule
        \rowcolor{grey}\multicolumn{6}{c}{General-purpose LLMs}\\
        \midrule
                Davinci & 0.631 & 0.834  & 0.634& 0.581 & 0.606 \\
                ChatGPT & 0.651  & 0.835  & 0.693 & 0.536 & 0.604 \\
                Llama2 & 0.595 & 0.799& 0.556& 0.608&0.581 \\
        \midrule
        \rowcolor{grey}\multicolumn{6}{c}{Logging-specific LLMs}\\
        \midrule
                LANCE$^\dagger$ & 0.612 & 0.822  & 0.667 & 0.420 & 0.515 \\
        \midrule
        \rowcolor{grey}\multicolumn{6}{c}{Code-based LLMs}\\
        \midrule
                InCoder & 0.608 & 0.800  & 0.712 & 0.655& 0.682 \\
                CodeGeex & 0.673 & 0.855  & 0.704 & 0.616 & 0.657 \\
                TabNine & 0.734 & 0.880 & 0.729 & 0.670 & 0.698\\
                Copilot & \textbf{\underline{0.743}}  & \textbf{\underline{0.882}}  & 0.722 & \textbf{\underline{0.703}} & 0.712 \\
                CodeWhisperer & 0.741 & 0.881  & \textbf{\underline{0.787}} & 0.668 & \textbf{\underline{0.723}}\\

                CodeLlama & 0.614 & 0.814 & 0.583 & 0.603 & 0.593 \\
                StarCoder & 0.661 & 0.829 & 0.656 & 0.649 & 0.653 \\
            \bottomrule
            % \footnotesize
            \multicolumn{6}{l}{\parbox{\linewidth}{\footnotesize $^\dagger$ Since LANCE decides logging point and logging statements simultaneously, we only consider its generated logging statements with correct locations.}}
             \end{NiceTabular}%
     }
     \vspace{-0.1in}
\end{table}

\begin{table*}[htbp]
\small
    \centering
    \vspace{-0.1in}
    \caption{The effectiveness of LLMs in producing logging texts.}
        \vspace{-0.1in}
    \label{tab:rq1-result-text}
    \resizebox{0.8\linewidth}{!}{%
        \begin{NiceTabular}{l||ccccccc}
        \CodeBefore
        \Body
            \toprule
                \Block{3-1}{\textbf{Model}} &
                \Block[c]{1-7}{\textbf{Logging Texts}} &
                & & &
                \\
            \cmidrule{2-8}
                &
                \textbf{BLEU-1} &
                \textbf{BLEU-2} &
                \textbf{BLEU-4} &
                \textbf{ROUGE-1} &
                \textbf{ROUGE-2} &
                \textbf{ROUGE-L} &
                \textbf{Semantics Similarity} \\
            \midrule
        \rowcolor{grey}\multicolumn{8}{c}{General-purpose LLMs}\\
        \midrule
                Davinci & 0.288 & 0.211  & 0.138 & 0.295 & 0.127 & 0.286 & 0.617\\
                ChatGPT & 0.291 & 0.217  & 0.149 & 0.306 & 0.142 & 0.298 & 0.633\\
                Llama2 & 0.235 &0.168 &0.102 &0.264 & 0.116& 0.261&0.569 \\
        \midrule
        \rowcolor{grey}\multicolumn{8}{c}{Logging-specific LLMs}\\
        \midrule
                LANCE$^\dagger$ & 0.306 & 0.236  & 0.167 & 0.162 & 0.078 & 0.162 & 0.347\\
        \midrule
        \rowcolor{grey}\multicolumn{8}{c}{Code-based LLMs}\\
        \midrule
                InCoder & 0.369 & 0.288  & 0.203 & 0.390 & 0.204 & 0.383 & 0.640\\
                CodeGeex & 0.330 & 0.248  & 0.160 & 0.339 & 0.149 & 0.333 & 0.598\\
                TabNine & 0.406 & 0.329 & 0.242 & 0.421 & 0.241 & 0.415 & 0.669\\
                Copilot & \textbf{\underline{0.417}} & \textbf{\underline{0.338}}  & 0.244 & \textbf{\underline{0.435}} & 0.247 & \textbf{\underline{0.428}} &  \textbf{\underline{0.703}}\\
                CodeWhisperer & 0.415 & 0.338 & \textbf{\underline{0.249}} & 0.430 & \textbf{\underline{0.248}} & 0.425 & 0.672\\
                CodeLlama & 0.216 & 0.146 & 0.089 & 0.258& 0.103& 0.251& 0.546 \\
                StarCoder & 0.353 & 0.278 & 0.195 & 0.378 &0.195 & 0.369 & 0.593\\
            \bottomrule
            % \footnotesize
            \multicolumn{8}{l}{\parbox{0.8\linewidth}{\footnotesize $^\dagger$ Since LANCE decides logging point and logging statements simultaneously, we only consider its generated logging statements with correct locations.}}
             \end{NiceTabular}%
     }
\end{table*}

(1) Logging levels.
Following previous studies~\cite{li2021deeplv,liu2022tell}, we use the level accuracy \textit{(L-ACC)} and Average Ordinal Distance Score \textit{(AOD)} for evaluating logging level predictions.
L-ACC measures the percentage of correctly predicted log levels out of all suggested results.
AOD~\cite{li2021deeplv} considers the distance between logging levels. Consequently, given the five logging levels in their severity order, i.e., \fixedwidth{error, warn, info, debug, trace}, the distance of $Dis(error, warn)=1$ is shorter than the distance of $Dis(error, info) = 2$.
% the \fixedwidth{error} is closer to \fixedwidth{warn} compared with \fixedwidth{info} and computes the distance between logging levels, since different levels are not independent of each other. 
% For example, the \fixedwidth{error} is closer to \fixedwidth{warn} compared with \fixedwidth{info}. 
AOD takes the average distance between the actual logging level $a_i$ and the suggested logging level (denoted as $Dis(a_i,s_i)$). AOD is therefore formulated as $AOD=\frac{\sum^N_{i=1} (1-Dis(a_i,s_i)/MaxDis(a_i))}{N}$, where $N$ is the number of logging statements and $MaxDis(a_i)$ refers to the maximum possible distance of the actual log level. 
% For instance, the maximum distance of \fixedwidth{error} is ~\textit{4} from \fixedwidth{trace} if there are ~\textit{5} levels (i.e., trace, debug, info, warn, error).

(2) Logging variables.
Evaluating predictions from LLMs is different from neural-based classification networks, as the predicted probabilities of each variable are not known. We thus employ \textit{Precision}, \textit{Recall}, and \textit{F1} to evaluate predicted logging variables. 
For each predicted logging statement, we use $S_{pd}$ to denote variables in LLM predictions and $S_{gt}$ to denote the variables in the actual logging statement. We report the proportion of correctly predicted variables (precision=$\frac{S_{pd} \cap S_{gt}}{S_{pd}}$), the proportion of actual variables predicted by the model (recall=$\frac{S_{pd} \cap S_{gt}}{S_{gt}}$), and their harmonic mean (F1=$2*\frac{Precision*Recall}{Precision+Recall}$).

(3) Logging texts. 
To align with previous research~\cite{mastropaolo2022using,ding2022logentext}, we assess the quality of the produced logging texts using two well-established machine translation evaluation metrics: \textit{BLEU}~\cite{papineni2002bleu} and \textit{ROUGE}~\cite{lin2004rouge}. 
These n-gram metrics compute the similarity between generated log messages and the actual logging text crafted by developers, yielding a percentage score ranging from \textit{0} to \textit{1}. A higher score indicates greater similarity between the generated log messages and the actual logging text. In particular, we use BLEU-K ($K=\{1,2,4\}$) and ROUGE-K ($K=\{1,2,L\}$) to compare the overlap concerning K-grams between the generated and the actual logs. In addition to the token-based match in a sparse space, we also incorporate \textit{semantic similarity} in our evaluation. Following prior works~\cite{gao2023constructing,ding2023crosscodeeval,xu2023prompting}, we also leverage widely-used code embedding models, UniXcoder~\cite{guo2022unixcoder} and OpenAI embedding~\cite{openaiemb}, to embed the logging texts to calculate the semantics similarity between generated and original logging texts, offering another evaluation metric from a semantic perspective.

% \ry{why choose BLEU124 but missing 3?} \zh{one reviewer of ASE also thinks BLUE3 is missing}

\subsection{RQ1: How do different LLMs perform in deciding ingredients of logging statements generation?}\label{sec:rq1}

% To help developers write appropriate logging statements, previous work developed models to automatically decide the log level or suggest log variables.
% Following the literature~\cite{he2021survey}, we identify three key ingredients of an effective logging statement based on the composition of logging statements:
% (1) \textit{Logging levels}~\cite{li2021deeplv,liu2022tell}: indicates the severity of a log event, helping developers to filter logs during system maintenance;
% (2) \textit{Logging variables}~\cite{liu2019variables,lidid,li2017log}: contains essential run-time information from system states; and
% (3) \textit{Logging texts}~\cite{mastropaolo2022using,ding2022logentext,lal2016logoptplus}: provides a description of the system activities and has been most frequently used for system maintenance. 
% Taking \fixedwidth{log.warn("Failed to connect to host: " + URL)} as an example, \fixedwidth{warn} and \fixedwidth{URL} refer to logging level and logging variables, respectively, where logging text is \fixedwidth{Failed to connect to host: <*>}.
% Although different ingredients offer diverse information for system maintainers, they serve as indispensable resources for reasoning the software behavior. 
% we evaluate the LLMs in \Odata
% By evaluating these critical ingredients, we provide a detailed analysis of the strengths and weaknesses of LLMs in producing complete and informative logging statements for developers. 
% \subsubsection{Settings}

% \subsubsection{Results and analysis}

To answer RQ1, we evaluate eleven top-performing LLMs on the benchmark dataset \Odata. The evaluation results are shown in Table~\ref{tab:rq1-result-level-variable} (levels, variables) and Table~\ref{tab:rq1-result-text} (logging texts), where we \textbf{\underline{underline}} the best performance score for each metric.

%%% 1. 分开说，logging level上copilot好，variables里codewhisper稍好，logging texts里...好。
\textbf{Intra-ingredient.} Regarding the logging levels, we observe that Copilot achieves the best L-ACC performance, i.e.,~\textit{0.743}, indicating that it can accurately predict 74.3\% of the logging levels.
While other baselines do not perform as well as Copilot, they also accurately suggest logging levels for at least \textit{60\%} logging statements. 
Compared with logging levels, there are greater differences among models when recommending logging variables. While~\textit{70\%} of the variables are recommended by Copilot, LANCE can only correctly infer~\textit{42\%} of them. 
% While CoderWhisper recommends variables with an F1 score of 0.723, LANCE only reaches the F1 score of 0.515. 
The recall rate for variable prediction is consistently lower than the precision rate across models, indicating the difficulty of identifying many of the variables. 
Predicting variables is more challenging than logging levels, as variables are diverse, customized, and have different meanings across systems. To address this challenge, logging variables should be inferred based on a deeper comprehension of code structure, such as control flow information.  

Concerning logging text generation shown in Table~\ref{tab:rq1-result-text}, both Copilot and CodeWhisperer demonstrate comparable performance across syntax-based metrics (BLEU, ROUGE) and semantic-based metrics, outperforming other baselines by a wide margin.
The comparison between syntax-based metrics and semantic-encoding metrics reveals a consistent trend across various LLMs: models exhibiting strong syntax similarity also exhibit high semantic similarity.  
On average, the studied models produce logging statements with a similarity of \textit{0.194} and \textit{0.341} for BLEU-4 and ROUGE-L scores, respectively. 
% \yc{AND semantics similarity?}
The result indicates that recommending appropriate logging statements remains a great challenge.

% By further comparing the syntax-based metrics and semantic-encoding metrics, Table~\ref{tab:rq1-result-level-variable} reveals that they 

% 

\begin{tcolorbox}[boxsep=1pt,left=2pt,right=2pt,top=3pt,bottom=2pt,width=\linewidth,colback=white!90!black,boxrule=0pt, colbacktitle=white!,toptitle=2pt,bottomtitle=1pt,opacitybacktitle=0]
\textbf{Finding 1.} \textit{While existing models correctly predict levels for~\textit{74.3\%} of logging statements, there is significant room for improvement in producing logging variables and logging texts.
% they only produce logging texts that are~\textit{24.9\%} similar to the actual ones based on BLEU-4, indicating that their logging statement generation abilities warrant improvements.
}
\end{tcolorbox}

%%% 2. 交叉看，趋势大致相同，incoder例外。
\textbf{Inter-ingredient.}
%%% 
From the inter-ingredient perspective, we observe that LLM performance trends are \textit{not consistently the same} across various ingredients, e.g., models that perform well in logging level prediction do not necessarily excel in generating logging texts. 
For instance, Incoder fares worst in predicting logging levels but performs better in generating logging texts (the fourth best performer). Upon manual investigation, we observe that Incoder predicts~\textit{41\%} of the cases with a \fixedwidth{debug} level, most of which are actually intended for the \fixedwidth{info} level statements.
Nevertheless, either Copilot or CodeWhisperer outperforms other baselines in all reported metrics.
This is likely because suggesting the three ingredients requires similar code comprehension capabilities, such as understanding data flows, specific code structures, and inferring code functionalities.

\begin{tcolorbox}[boxsep=1pt,left=2pt,right=2pt,top=3pt,bottom=2pt,width=\linewidth,colback=white!90!black,boxrule=0pt, colbacktitle=white!,toptitle=2pt,bottomtitle=1pt,opacitybacktitle=0]
\textbf{Finding 2.} \textit{LLMs may perform inconsistently on deciding different ingredients, making model comparisons more difficult based on multiple ingredient-wise metrics.}
\end{tcolorbox}

\subsection{RQ2: How do LLMs compare to conventional logging models in logging ability?}\label{sec:rq2_1}

 \begin{figure*}[tbp]
  \centering
     \includegraphics[width=\linewidth]{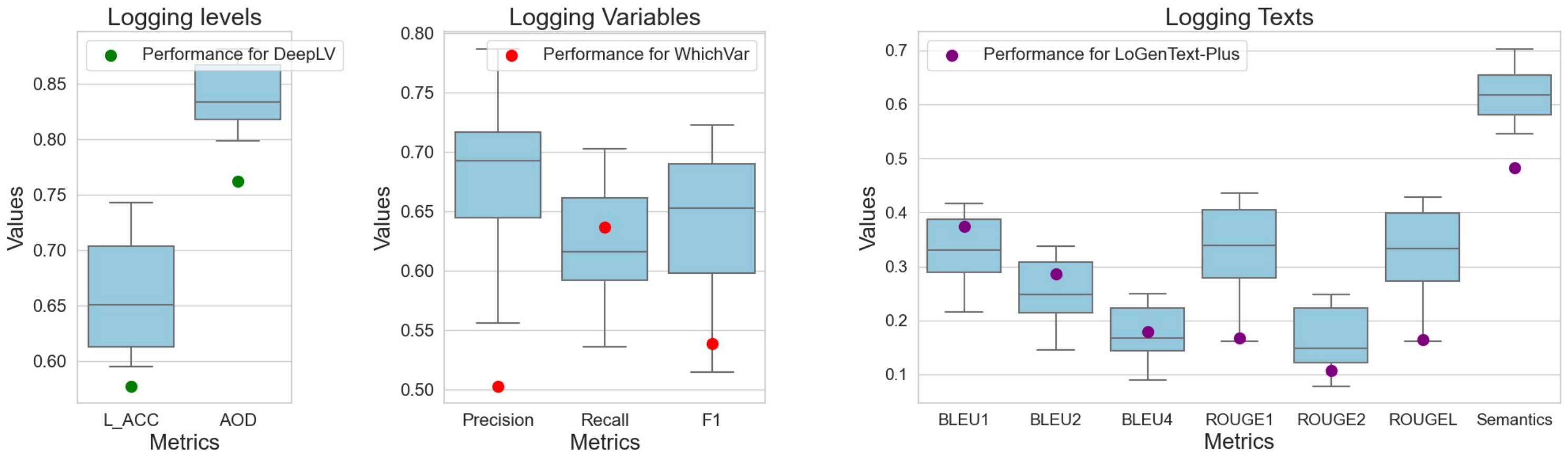}
     \caption{Comparison between traditional logging models and LLM-powered models.}
     \label{fig:rq2-traditional}
 \end{figure*}

We compare the results of directly using LLMs for logging against conventional logging models on \Odata. As conventional logging models can only predict one ingredient, we opt for state-of-the-art models for each one (i.e., DeepLV, WhichVar, and LoGenText-Plus) and present their performance against LLMs in Fig.~\ref{fig:rq2-traditional}. The boxplot illustrates the performance range of LLM-powered models, while the points depict conventional logging models.

  \begin{figure}[tbp]
  \centering
     \includegraphics[width=\columnwidth]{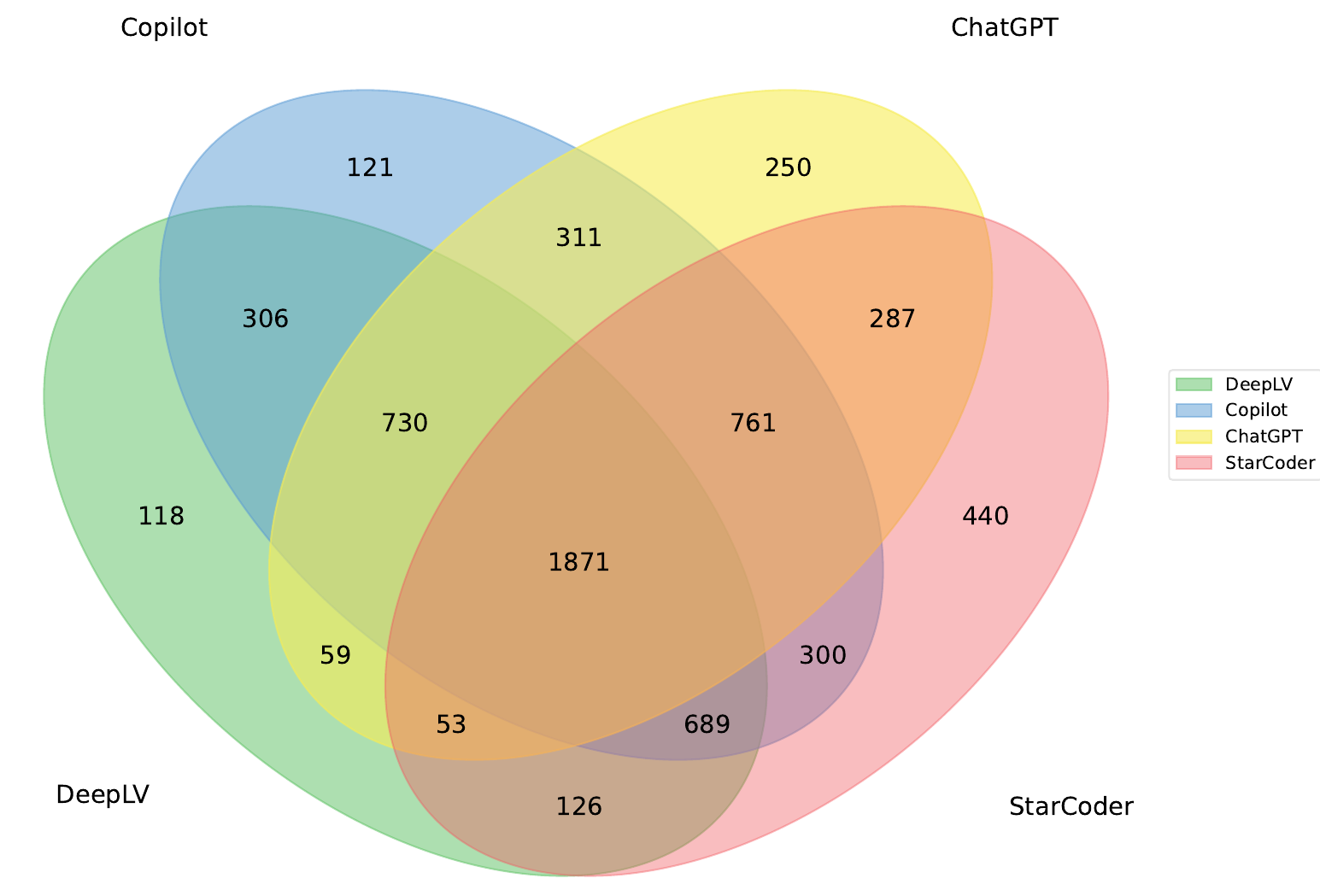}
     \caption{Venn diagram for logging levels prediction.}
     \label{fig:rq2-venn}
 \end{figure}

Despite being carefully designed for the logging task, the conventional logging models do not surpass LLMs. As shown in Fig.~\ref{fig:rq2-traditional}, conventional models exhibit inferior performance compared to any LLMs on five metrics (i.e., below the lower whiskers) and fall below the median on the other three metrics (i.e., below the line in the box).
In terms of logging level prediction, DeepLV performs worse than any of our studied LLMs, correctly predicting only 57.7\% of statements. Regarding generating logging variables and texts, WhichVar and LoGenText-Plus show comparable performance to LANCE, but lag behind other studied LLMs. While the most effective model (Copilot) achieves a 0.703 semantic-based similarity in logging texts, the state-of-the-art logging model, LoGenText-Plus, only produces a 0.485 similarity (yielding a 21.8\% drop).
These surprising results show that, \textit{without any specific change or fine-tuning, directly applying LLMs for logging statement generation yields better performance compared to conventional logging baselines.}

  \begin{figure}[tbp]
  \centering
     \includegraphics[width=\columnwidth]{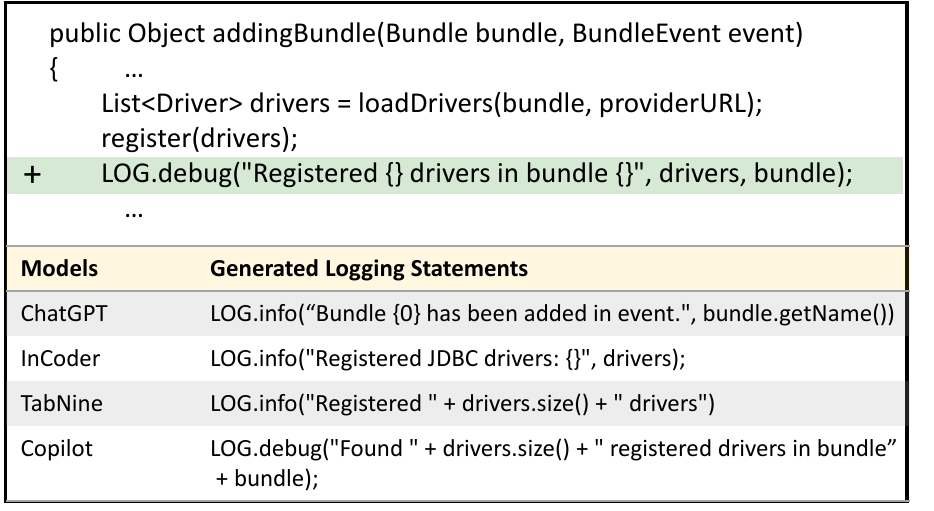}
     \vspace{-0.2in}
     \caption{An example of the generation results from eight models.}
\vspace{-0.2in}
     \label{fig:rq2-case}
 \end{figure}

Figure~\ref{fig:rq2-venn} displays the Venn diagram illustrating the logging levels correctly predicted by DeepLV in comparison to three chosen LLMs on the \Odata dataset.
Notably, 97\% of the cases handled by DeepLV can also be predicted by LLMs. In contrast, DeepLV can only handle 70\%, 62\%, and 60\% of the cases successfully predicted by Copilot, ChatGPT, and StarCoder, respectively. 
% The overlapping areas further underscore the promising potential of LLMs for logging practices.

To demonstrate the ability of LLMs, we present Fig. ~\ref{fig:rq2-case} to illustrate some statements produced by ChatGPT, InCoder, Copilot, and TabNine, respectively. Through pre-training, these LLMs gain a basic understanding of method activity in adding bundles with drivers, leading to the generation of relevant logging variables.
Notably, code-based LLMs produce more accurate logging statements compared to models pre-trained for general purposes. In Fig. ~\ref{fig:rq2-case}, general-purpose LLMs (i.e., ChatGPT) mispredict the logging statement by focusing on the \textcolor{red}{\fixedwidth{event}} variable in the method declaration, overlooking the driver registration process preceding the logging point. 
Conversely, most code models (e.g., InCoder) capture such processes, recognizing that \textcolor{green}{\fixedwidth{drivers}} are critical variables describing a device status. We attribute the performance difference to the gap between natural and programming languages. Training on a code base enables these models to acquire programming knowledge, bridging the gap and enhancing logging performance.

\begin{tcolorbox}[boxsep=1pt,left=2pt,right=2pt,top=3pt,bottom=2pt,width=\linewidth,colback=white!90!black,boxrule=0pt, colbacktitle=white!,toptitle=2pt,bottomtitle=1pt,opacitybacktitle=0]
\textbf{Finding 3.} \textit{When directly applying LLMs to logging statement generation, without fine-tuning, they still yield better performance than conventional logging baselines.}
\end{tcolorbox}

\subsection{RQ3: How do the prompts for LLMs affect logging performance?}\label{sec:rq3}
Previous literature has identified the variance of input prompts can significantly affect the performance of LLMs'~\cite{gao2023constructing}. For the LLMs that can take prompts (e.g., ChatGPT, LLaMa2), we investigate the influences of instructions and demonstrate examples for their logging purpose.

\textbf{Impact of different instructions.}
LLMs have been shown to be sensitive to the instructions used to query the LLM sometimes. 
To compare the impact of different instructions, we conducted a two-round survey involving 54 developers from a world-leading technical company, each possessing a minimum of two years of development experience. 
To begin with, we ask the developers to individually propose 10 instructions that they would consider when utilizing LLMs for generating logging statements. Subsequently, we distributed a second questionnaire, asking developers to choose the top 5 instructions from the initial round that they likely to employ. Eventually, instructions receiving the top 5 votes will be considered for evaluation, shown as follows.
\begin{enumerate}[leftmargin=*]
    \item Your task is to generate the logging statement for the corresponding position.
    \item You are an expert in software DevOps; please help me write the informative logging statement.
    \item Complete the logging statement while taking the surrounding code into consideration.
    \item Your task is to write the corresponding logging statement. Note that you should keep consistent with current logging styles.
    \item Please help me write an appropriate logging statement below.
\end{enumerate}

\begin{figure}[tbp]
  \centering
     \includegraphics[width=\columnwidth]{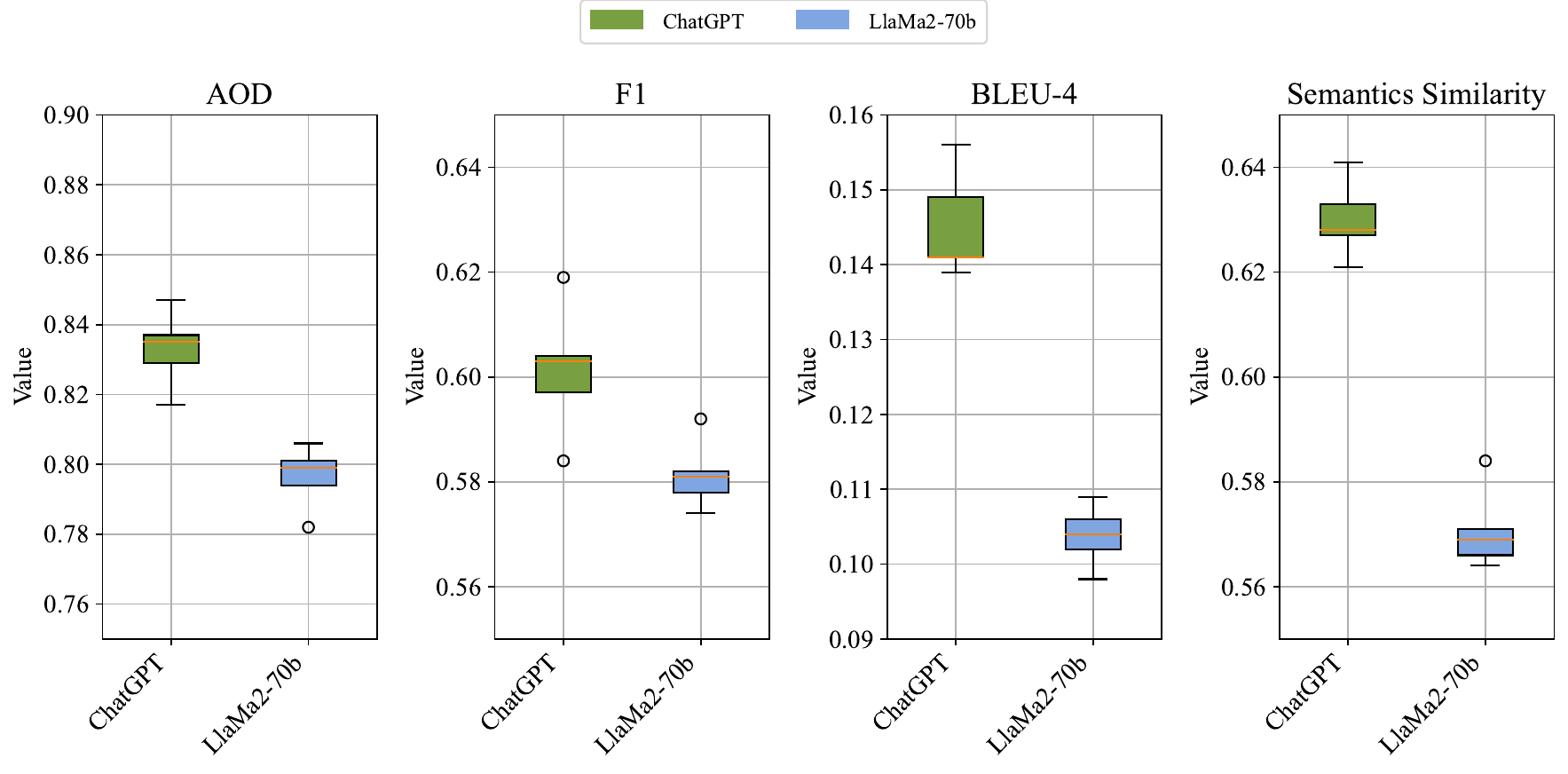}
     \caption{The selected metrics of LLMs’ logging performance with different instructions.}
     \label{fig:instruction}
 \end{figure}

We then feed these representative instructions into two studied LLMs, that is, ChatGPT, and LLaMa2, respectively. 
The box plot in Fig.~\ref{fig:instruction} exhibits logging performance associated with different instructions.
The selected instructions result in approximately 3\% performance variance for each metric, revealing the importance of designing prompts. Among all metrics, the difference in logging variable prediction for ChatGPT is slightly larger, but still in the range of 4\% variation.
Despite there being small variations due to different instructions, these variances do not alter the consistent superiority of ChatGPT over LLaMa2. In summary, as long as the logging ability of LLMs is evaluated using the same instructions, such evaluation and comparison are meaningful.

\begin{tcolorbox}[boxsep=1pt,left=2pt,right=2pt,top=3pt,bottom=2pt,width=\linewidth,colback=white!90!black,boxrule=0pt, colbacktitle=white!,toptitle=2pt,bottomtitle=1pt,opacitybacktitle=0]
\textbf{Finding 4.} \textit{Although instructions influence LLMs to varying extents, there is cohesiveness in the relative ranking of LLMs with the same instructions.}
\end{tcolorbox}

% cohesiveness of chatgpt, llama2 trend,  the relative ranking
% can variance, but not affect our exps.
 
% the collected instructions among them to determine the top 5 instructions deemed most suitable, optimal, and likely to be used by the group.
% distributed a questionnaire to 54 developers in a world-leading technical company, each with a minimum of two years of experience in development. In the questionnaire, we ask the developers to individually propose \yt{how many?} instruction they would consider using if they were to employ LLMs for creating logging statements. 
% Upon their response, we distribute 
% After we collected their responses, we conducted a vote among them to determine the top 5 instructions deemed most suitable, optimal, and likely to be used by the group.

\begin{figure*}[tbp]
    \centering
    \includegraphics[width=\linewidth]{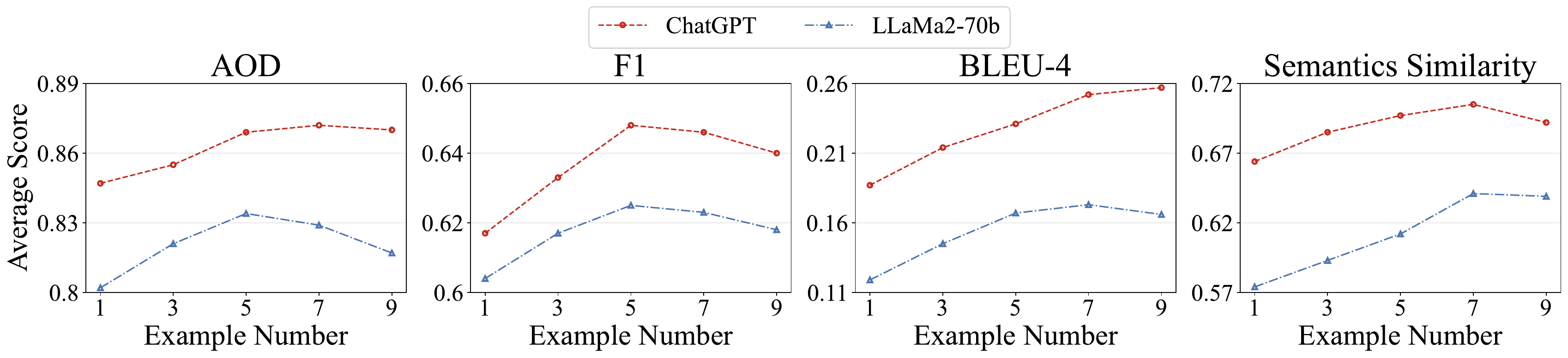}
    \vspace{-0.2in}
    \caption{The selected metrics of LLMs' logging performance with different numbers of examples.}
    \label{fig:icl}
\end{figure*}

% We evaluated their instructions

\textbf{Impact of different numbers of logging examples.} 
In-context learning (ICL) is a prevalent prompt strategy, enabling LLMs to glean insights from few-shot examples in the context. Many studies have shown that LLMs can boost complicated code intelligence tasks through ICL implementation \cite{gao2023constructing}. Despite being promising, there are intriguing properties that require further exploration, for example, the effects of parameter settings in ICL.

Fig. \ref{fig:icl} presents the logging performance (i.e., logging level, variable, texts) in terms of different numbers of demonstration examples provided. In this experiment, we vary the number of demonstrations for ChatGPT and LLaMa2 from 1 to 9. We select and order demonstration examples measured by using BM25 retrieval methods, as previous works have demonstrated its effectiveness in code tasks~\cite{gao2023constructing}.

%\yt{BM-25?} similarity.

The figure illustrates the impact of the number of demonstration examples on LLMs' logging performance, resulting in a increment of 2\%-8\%.
Initially, the performance of ICL improves across all metrics as the number of demonstration examples increases. However, when the number of examples surpasses 5, divergent trends emerge for different tasks. For instance, in determining logging levels (AOD) and logging variables (F1), the LLaMa performance peaks at 5 demonstration examples but experiences a decline with further increments to 7. Conversely, in logging text generation (BLEU-4, Semantics Similarity), LLaMa performance continues to rise and stabilizes beyond 7 examples. We attribute these diverse trends to the \textit{model distraction problem}~\cite{yuan2023evaluating}. Tasks involving predicting logging levels and variables demand an intricate analysis of individual program structures and variable flows, and the introduction of additional examples with longer input lengths can potentially distract the model, leading to performance degradation. In contrast, logging text generation involves a high-level program understanding and summarization. More examples allow LLMs to learn proper logging styles from other demonstrations.

%\yt{how many examples in the baseline? rmb to mention in implementation}

\begin{tcolorbox}[boxsep=1pt,left=2pt,right=2pt,top=3pt,bottom=2pt,width=\linewidth,colback=white!90!black,boxrule=0pt, colbacktitle=white!,toptitle=2pt,bottomtitle=1pt,opacitybacktitle=0]
\textbf{Finding 5.} \textit{More demonstration examples in the prompt do not always improve performance. It is recommended to use 5-7 examples in the demonstration to achieve optimal results.}
\end{tcolorbox}

\subsection{RQ4: How do external factors influence the effectiveness in generating logging statements?}\label{sec:rq4}

\begin{table*}[htbp]
    \centering
     \caption{The results of logging statement generation without comments.}
    \label{tab:rq4-without-comment}
    \resizebox{0.85\linewidth}{!}{%
        \begin{NiceTabular}{l||c|c|ccc}
        \CodeBefore
        \Body
            \toprule
                \Block{3-1}{\textbf{Model}} &
                \textbf{Logging Levels} & \textbf{Logging Variables} &
                \Block[c]{1-3}{\textbf{Logging Texts}} &
                &
                \\
            \cmidrule{2-6}
                & 
                \textbf{AOD} &
                \textbf{F1} &
                \textbf{BLEU-4} &
                \textbf{ROUGE-L} &
                \textbf{Semantics Similarity} \\
            \midrule
            Davinci & 0.834 (0.0\%-) & 0.587 (3.1\%$\downarrow$) & 0.133 (3.6\%$\downarrow$) & 0.283 (1.0\%$\downarrow$) & 0.608 (1.5\%$\downarrow$)\\
            ChatGPT & 0.833 (0.2\%$\downarrow$) & 0.592 (2.0\%$\downarrow$) & 0.149 (0.0\%-) & 0.294 (1.3\%$\downarrow$) & 0.614 (3.0\%$\downarrow$)\\
            Llama2 & 0.789 (1.3\%$\downarrow$) & 0.574 (1.2\%$\downarrow$) & 0.099 (2.9\%$\downarrow$) & 0.255 (2.3\%$\downarrow$) & 0.544 (4.4\%$\downarrow$)\\
            InCoder & 0.789 (1.4\%$\downarrow$) &  0.674 (1.2\%$\downarrow$)  & 0.201 (1.0\%$\downarrow$)& 0.377 (9.2\%$\downarrow$) &  0.622 (2.8\%$\downarrow$)\\
            CodeGeex & 0.848 (0.8\%$\downarrow$) & 0.617 (6.1\%$\downarrow$) & 0.149 (6.9\%$\downarrow$) & 0.306 (8.1\%$\downarrow$) & 0.578 (3.3\%$\downarrow$)\\
            TabNine & 0.876 (0.5\%$\downarrow$) & 0.690 (1.1\%$\uparrow$) & 0.239 (1.2\%$\downarrow$) & 0.412 (0.7\%$\downarrow$) & 0.655 (2.1\%$\downarrow$)\\
            Copilot &\textbf{\underline{0.878}} (0.5\%$\downarrow$) & 0.696 (2.2\%$\downarrow$) & 0.241 (1.2\%$\downarrow$) & \textbf{\underline{0.419}} (2.1\%$\downarrow$)  & \textbf{\underline{0.689}} (2.0\%$\downarrow$)\\
            CodeWhisperer & 0.877 (0.7\%$\downarrow$)& \textbf{\underline{0.718}} (0.7\%$\downarrow$) & \textbf{\underline{0.244}} (2.0\%$\downarrow$)& 0.418 (1.6\%$\downarrow$)  & 0.661 (1.6\%$\downarrow$)\\
            CodeLlama & 0.804 (1.2\%$\downarrow$) & 0.581 (2.0\%$\downarrow$) &  0.087 (2.2\%$\downarrow$) & 0.247 (1.6\%$\downarrow$) & 0.544 (0.3\%$\downarrow$)\\
            StarCoder & 0.823 (0.7\%$\downarrow$)& 0.647 (0.9\%$\downarrow$)& 0.193 (1.0\%$\downarrow$)& 0.369 (2.4\%$\downarrow$)& 0.591 (\textbf{\underline{0.3\%$\downarrow$}})\\
            \midrule
            Avg.$\Delta$ & 0.835 (0.8\%$\downarrow$) & 0.638 (2.1\%$\downarrow$) & 0.173 (2.2\%$\downarrow$)  & 0.338 (3.0\%$\downarrow$) & 2.1\%$\downarrow$\\
            \bottomrule
            % \multicolumn{6}{l}{\parbox{\linewidth}{\footnotesize $^\dagger$ Since LANCE decides logging point and logging statements simultaneously, we only consider its generated logging statements with correct locations.}}
             \end{NiceTabular}%
     }
\end{table*}

While RQ3 discusses the prompt construction for LLMs, some external program information is likely to affect their effectiveness in logging generation. In particular, we focus on how \textit{comments} and \textit{the scope of programming contexts} will impact the model performance. 
% With the investigation, not only can developers make the best use of these tools, but also they can guide researchers in building more effective automated logging models.
% In particular, we focus on how the different \textit{input range} and \textit{comments} will impact the performance.

  \begin{figure}[tbp]
   \centering
     \includegraphics[width=0.9\columnwidth]{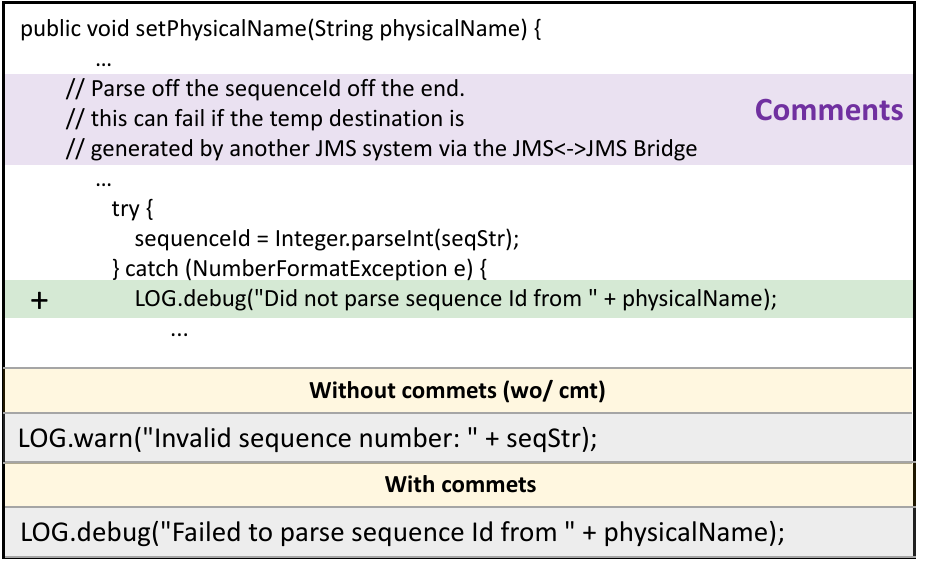}
     \vspace{-0.1in}
     \caption{A logging statement generation case using code comments.}
     \vspace{-0.2in}
     \label{fig:rq3-comment-case}
 \end{figure}

\begin{table*}[htbp]
    \centering
    \caption{The results of logging statement generation with file-evel contexts.}
    \label{tab:rq4-with-file}
    \resizebox{0.85\linewidth}{!}{%
        \begin{NiceTabular}{l||c|c|ccc}
        \CodeBefore
        \Body
            \toprule
                \Block{3-1}{\textbf{Model}} &
                \textbf{Logging Levels} & \textbf{Logging Variables} &
                \Block[c]{1-3}{\textbf{Logging Texts}} &
                &
                \\
            \cmidrule{2-6}
                & 
                \textbf{AOD} &
                \textbf{F1} &
                \textbf{BLEU-4} &
                \textbf{ROUGE-L} &
                \textbf{Semantics Similarity} \\
            \midrule
            Davinci & 0.854 (2.6\%$\uparrow$) & 0.638 (5.3\%$\uparrow$) & 0.156 (13.0\%$\uparrow$) & 0.318 (11.2\%$\uparrow$) & 0.635 (2.9\%$\uparrow$)\\
            ChatGPT & 0.858 (2.8\%$\uparrow$) & 0.650 (7.6\%$\uparrow$) & 0.253 (51.5\%$\uparrow$) & 0.389 (30.5\%$\uparrow$) & 0.704 (11.2\%$\uparrow$)\\
            Llama2 & 0.832 (4.1\%$\uparrow$) & 0.617 (6.2\%$\uparrow$) & 0.149 (46.1\%$\uparrow$)& 0.392 (50.2\%$\uparrow$)& 0.669 (17.6\%$\uparrow$)\\
            InCoder & 0.815 (1.9\%$\uparrow$) & 0.745 (9.2\%$\uparrow$) & 0.307 (51.2\%$\uparrow$) & 0.521 (35.3\%$\uparrow$) & 0.734 (11.7\%$\uparrow$)\\
            CodeGeex & 0.869 (1.6\%$\uparrow$) & 0.696 (5.9\%$\uparrow$) & 0.241 (50.6\%$\downarrow$) & 0.395 (18.6\%$\uparrow$) & 0.644 (7.7\%$\uparrow$)\\
            TabNine & 0.912 (3.6\%$\uparrow$) & 0.767 (9.9\%$\uparrow$) & 0.375 (55.0\%$\uparrow$) & 0.530 (27.7\%$\uparrow$) & 0.783 (17.0\%$\uparrow$)\\
            Copilot & \textbf{\underline{0.916}} (3.9\%$\uparrow$) & 0.742 (4.2\%$\uparrow$) & 0.346 (41.8\%$\uparrow$) & 0.522 (22.0\%$\uparrow$) & \textbf{\underline{0.816}} (16.1\%$\uparrow$)\\
            CodeWhisperer & 0.913 (3.6\%$\uparrow$) & \textbf{\underline{0.792}} (9.6\%$\uparrow$) & \textbf{\underline{0.401}} (61.0\%$\uparrow$) & \textbf{\underline{0.559}} (31.5\%$\uparrow$) & 0.811 (20.7\%$\uparrow$)\\
            CodeLlama & 0.817 (0.4\%$\uparrow$) & 0.607 (2.4\%$\uparrow$)& 0.144 (61.8\%$\uparrow$) & 0.378 (50.6\%$\uparrow$) & 0.642 (17.6\%$\uparrow$)\\
            StarCoder & 0.847 (2.2\%$\uparrow$)& 0.714 (9.3\%$\uparrow$)& 0.314 (61.0\%$\uparrow$)& 0.517 (40.1\%$\uparrow$)& 0.679 (14.5\%$\uparrow$)\\
            \midrule
            Avg.$\Delta$ & 2.7\%$\uparrow$ & 6.9\%$\uparrow$ & 49.3\%$\uparrow$ & 31.8\%$\uparrow$& 13.7\%$\uparrow$\\
            \bottomrule
            % \multicolumn{6}{l}{\parbox{\linewidth}{\footnotesize $^\dagger$ Since LANCE decides logging point and logging statements simultaneously, we only consider its generated logging statements with correct locations.}}
             \end{NiceTabular}%
     }
\end{table*}

\textbf{With comment v.s. without comment.}
Inspired by the importance of human-written comments for intelligent code analysis~\cite{guo2022unixcoder, mastropaolo2023robustness, wan2018improving}, we also explore the utility of comments for logging.
% \yc{as some models neglect to embed comments as features~\cite{mastropaolo2022using,he2018characterizing,liu2022tell}}. 
To this end, we feed the original code (with comment) and comment-free code into LLMs separately, compare their results, and analyze the corresponding performance drop rate ($\Delta$) in Table~\ref{tab:rq4-without-comment} in terms of AOD, F1, BLEU, and ROUGE score.
The results show that LLMs consistently encounter performance drops without comments, with an average drop rate on \textit{0.8\%}, \textit{2.1\%}, \textit{2.2\%}, and \textit{3.0\%} for AOD, F1, BLEU-4, and ROUGE-L, respectively. 
The reason is that,
comments are used to describe the functionalities of the corresponding code, thus sharing similarities to logging practices that record system activities.

Fig.~\ref{fig:rq3-comment-case} presents an example with CodeWhisperer that can be facilitated by reading the comment of \fixedwidth{parse sequence Id}. 
Without the comment, CodeWhisperer only concentrates on the invalid sequence number but fails to involve parsing descriptions, which may further mislead maintainers on parsing failure diagnosis. Moreover, the comments highlight that the exception is a foreseeable and potentially common issue, which helps the LLMs in correctly selecting the log level, changing the logging level from \fixedwidth{warn} to \fixedwidth{debug}.

\begin{tcolorbox}[boxsep=1pt,left=2pt,right=2pt,top=3pt,bottom=2pt,width=\linewidth,colback=white!90!black,boxrule=0pt, colbacktitle=white!,toptitle=2pt,bottomtitle=1pt,opacitybacktitle=0]
\textbf{Finding 6.} \textit{Ignoring code comments impedes LLMs in generating logging statements, resulting in an average \textit{2.43\%} decrease when recommending logging texts.
}
\end{tcolorbox}

 \begin{figure}[tbp]
 \centering
     \includegraphics[width=\columnwidth]{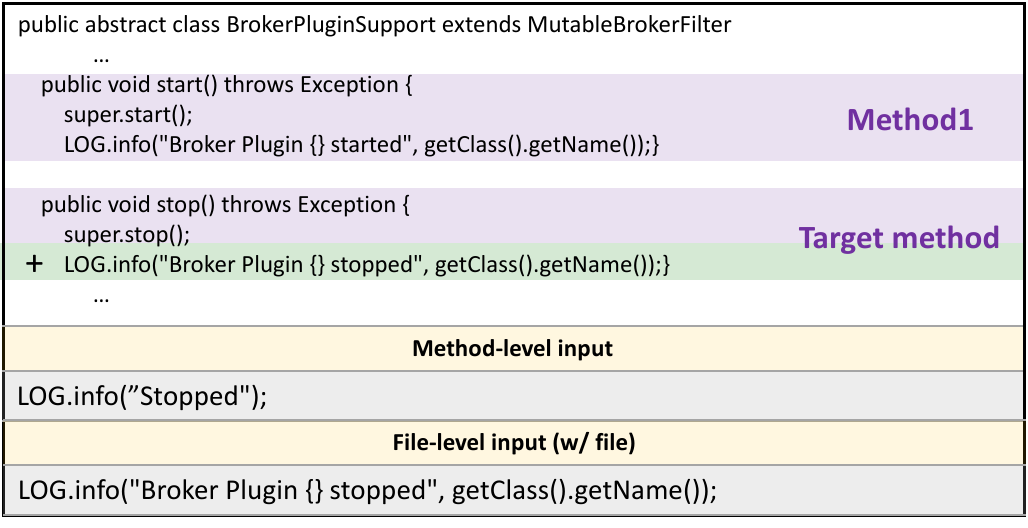}
     \vspace{-0.25in}
     \caption{A logging statement generation case using different programming contexts.}
     \label{fig:rq3-file-case}
          \vspace{-0.2in}
 \end{figure}

\textbf{Programming contexts: method v.s. file.}
Current logging practice tools restrict their work on code snippets or methods~\cite{mastropaolo2022using, ding2022logentext, liu2019variables}, and ignore the information from other related methods~\cite{dawes2023towards}.
However, methods that implement similar 
functionalities can contain similar logging statements~\cite{he2018characterizing}, which can be used as references to resolve logging statements.
In past works, this constraint was mainly due to the limits in input size in previous neural-based models.
But since LLMs can now process thousands of input tokens without suffering from such limitations, we aim to assess the benefits of larger programming contexts, i.e., file-level input.

In this regard, we feed \textit{an entire Java file} for generating logging statements rather than \textit{the target method}.
The result in Table~\ref{tab:rq4-with-file} presents the effectiveness of file-level input (w/ File) and the corresponding increment ratio ($\Delta$).
The result suggests that file-level programming contexts consistently enhance performance in terms of all metrics where, for example, TabNine increases~\textit{3.6\%},~\textit{9.9\%}, and~\textit{55.0\%} for AOD, F1, and BLEU score, respectively.
On average, all models generate logging statements that are \textit{49.3\%} more similar to actual ones (reflected by BLEU-4) than using a single method as input.
We take Fig.~\ref{fig:rq3-file-case} as an example from CodeWhisperer to illustrate how LLMs can learn from an additional method, where the green line represents the required logging statements. The model learned logging patterns from Method1, which includes the broker plugin name and its status (i.e., \fixedwidth{start}). Regarding \fixedwidth{stop()}, CodeWhisperer may refer to Method1 and write similar logging statements by changing the status from \fixedwidth{started} to \fixedwidth{stopped}. 
Additionally, by analyzing the file-level context, LLMs can identify pertinent variables, learn relationships between multiple methods, and recognize consistent logging styles within the file. 
% As a result, LLMs are able to generate more accurate and context-aware logs.
Last but not least, the comparison of Table~\ref{tab:rq4-without-comment} and Table~\ref{tab:rq4-with-file} implies that expanding the range of programming texts has a stronger impact than incorporating comments, even though certain models (e.g., Copilot) are trained to generate code from natural language.

\begin{tcolorbox}[boxsep=1pt,left=2pt,right=2pt,top=3pt,bottom=2pt,width=\linewidth,colback=white!90!black,boxrule=0pt, colbacktitle=white!,toptitle=2pt,bottomtitle=1pt,opacitybacktitle=0]
\textbf{Finding 7.} \textit{Compared to comments, incorporating file-level programming contexts leads to a greater improvement in logging practice 
by providing access to additional functionality-similar methods, variable definitions and intra-project logging styles.}
\end{tcolorbox}

% To analyze the impact of reception scope, we feed the entire code file for each model (instead of the only method) to predict the logging statements so that LLMs can utilize information from the external method.
% Besides, as previous studies present the importance of human-written comments in other intelligent code tasks\yt{cite}, we also explore the utility of comments for our logging study. To this end, we remove all comments from code and feed the comment-free code into LLMs to evaluate the effectiveness. 
% Experiment results from these influencing factors are shown in Table~\ref{tab:rq3-influencing-result}, which reports the representative metrics from three ingredients (i.e., AOD, F1, BLEU, and ROUGE), respectively.

\begin{table*}[htbp]
\small
    \centering
    \caption{The generalization ability of LLMs in producing logging statements for unseen code.}
    \label{tab:rq4-generalizability-result}
    \resizebox{0.9\linewidth}{!}{%
        \begin{NiceTabular}{l||cc|cc|cccccc|c}
        \CodeBefore
        \Body
            \toprule
                \Block{3-1}{\textbf{Model}} &
                \Block[c]{1-2}{\textbf{Levels}} & &
                \Block[c]{1-2}{\textbf{Variables}} & &
                \Block[c]{1-6}{\textbf{Texts}} & & & & & & 
                \textbf{Average}
                \\
            \cmidrule{2-12}
                &
                \textbf{AOD} &
                \textbf{$\Delta$} &
                \textbf{F1} &
                \textbf{$\Delta$} &
                \textbf{BLEU-4} &
                \textbf{$\Delta$} &
                \textbf{ROUGE-L} &
                \textbf{$\Delta$} &
                \textbf{Semantics} & 
                \textbf{$\Delta$} & 
                \textbf{Avg. $\Delta$}\\
            \midrule
        \rowcolor{grey}\multicolumn{12}{c}{General-purpose LLMs}\\
        \midrule
                Davinci & 0.820 & 1.7\%$\downarrow$ & 0.523 & 13.7\%$\downarrow$ & 0.116 & 15.9\%$\downarrow$ & 0.234 & 20.7\%$\downarrow$  & 0.533 & 13.6\%$\downarrow$ & 13.1\%$\downarrow$ \\
                ChatGPT & 0.830 & 0.6\%$\downarrow$  & 0.532 & 11.9\%$\downarrow$ & 0.118 & 20.8\%$\downarrow$ & 0.240 & 19.5\%$\downarrow$ & 0.541 & 14.5\%$\downarrow$ & 13.5\%$\downarrow$\\
                Llama2 & 0.788 & 1.4\%$\downarrow$& 0.568 & 2.2\%$\downarrow$& 0.094 &  7.8\%$\downarrow$& 0.213 & 18.4\%$\downarrow$ & 0.513 & 9.8\%$\downarrow$ & 7.9\%$\downarrow$\\
        \midrule
        \rowcolor{grey}\multicolumn{12}{c}{Logging-specific LLMs}\\
        \midrule
                LANCE & 0.817 & 0.6\%$\downarrow$ & 0.475 & 7.5\%$\downarrow$ & 0.153 & 8.4\%$\downarrow$  & 0.144 & \textbf{\underline{11.1\%$\downarrow$}} & 0.301 & 13.3\%$\downarrow$ & \textbf{\underline{8.2\%$\downarrow$}}\\
        \midrule
        \rowcolor{grey}\multicolumn{12}{c}{Code-based LLMs}\\
        \midrule
               InCoder & 0.778& 2.8\%$\downarrow$ & 0.587& 13.9\%$\downarrow$ & 0.175& 13.8\%$\downarrow$ & 0.316& 17.5\%$\downarrow$ & 0.584 &8.8\%$\downarrow$  & 11.4\%$\downarrow$\\
               CodeGeex & 0.850 & 0.6\%$\downarrow$ & 0.534 & 18.7\%$\downarrow$ & 0.115 & 28.1\%$\downarrow$ & 0.253 & 25.4\%$\downarrow$ &  0.549&  8.2\%$\downarrow$& 16.2\%$\downarrow$ \\
               TabNine & 0.869 & 1.3\%$\downarrow$ & 0.596 & 14.6\%$\downarrow$ & 0.202 & 16.5\%$\downarrow$ & 0.342 & 18.8\%$\downarrow$ & 0.608  & 9.1\%$\downarrow$ & 12.1\%$\downarrow$ \\
               Copilot & \textbf{\underline{0.881}} & \textbf{\underline{0.1\%$\downarrow$}} & 0.610 & 14.3\%$\downarrow$ & \textbf{\underline{0.234}} & \textbf{\underline{4.1\%$\downarrow$}} & \textbf{\underline{0.377}} & 13.3\%$\downarrow$ & \textbf{\underline{0.641}} & 8.8\%$\downarrow$ & 8.2\%$\downarrow$ \\
               CodeWhisperer & 0.871& 1.1\%$\downarrow$ & \textbf{\underline{0.629}} & 13.0\%$\downarrow$ & 0.219 & 12.0\%$\downarrow$ & 0.362 & 14.6\%$\downarrow$ & 0.612 & 8.9\%$\downarrow$&9.9\%$\downarrow$ \\
                % \textbf{Average} & \\
                CodeLlama  & 0.801 & 1.6\%$\downarrow$& 0.574 & \textbf{\underline{3.2\%$\downarrow$}} & 0.078 & 12.6\%$\downarrow$ & 0.211 & 15.9\%$\downarrow$ & 0.482 & 11.7\%$\downarrow$&9.0\%$\downarrow$ \\
                StarCoder  & 0.811 & 2.2\%$\downarrow$& 0.619 & 5.2\%$\downarrow$& 0.175 & 10.3\%$\downarrow$ & 0.309 & 16.3\%$\downarrow$ & 0.546 & \textbf{\underline{7.9\%$\downarrow$}}& 8.4\%$\downarrow$\\
        \midrule
        Avg. $\Delta$ & - & 1.4\%$\downarrow$ & - & 11.6\%$\downarrow$ & - & 15.0\%$\downarrow$ & - & 19.2\%$\downarrow$ & & 10.4\%$\downarrow$ & 11.5\%$\downarrow$\\
        \bottomrule
             \end{NiceTabular}%
     }
\end{table*}

\subsection{RQ5: How do LLMs perform in logging unseen code?}\label{sec:rq5}

In this RQ, we assess the generalization capabilities of language models by evaluating them on the \Tdata (Table~\ref{tab:codeTran}).
As stated in Section~\ref{sec:Tdata-creation}, predicting accurate logging statements does not necessarily imply that a model can be generalized to unseen cases well.
As the modern software codebase is continuously evolving, we must explore LLMs' ability to handle these unseen cases in daily development.
% LLM's ability to handle unseen cases in daily development is very crucial.

We present the result in Table~\ref{tab:rq4-generalizability-result}, where we \textbf{\underline{underline}} the best performance for each metric and the lowest performance drop rate ($\Delta$) compared to corresponding results in \Odata.
% the best performance of each metric and the lowest performance drop rate compared with evaluating on \Odata~ ($\Delta$) is \textbf{\underline{underlined}}. 
Our experiments show that all models experience different degrees of performance degradation when generating logging statements on unseen code.
LANCE has the smallest average decrease of~\textit{6.9\%} across metrics, while CodeGeex is most impacted with a \textit{16.2\%} drop.
Copilot exhibits the greatest generalization capabilities by outperforming other baselines for three out of four metrics on unseen code.
Additionally, we observe that predicting logging levels the smallest degradation in performance (\textit{1.4\%}), whereas
predicting logging variables and logging text (BLEU-4) experience significant performance drops,  \textit{11.6\%} and~\textit{15\%}, respectively.
Such experiments indicate that resolving logging variables and logging texts is more challenging than predicting logging levels, thus warranting more attention in future research.
% which require a comprehensive understanding of the context, 
% Nevertheless,
% We also observe that predicting logging levels has a minimal decline in the performance of 1.1\%, indicating 
% , indicating that our code transformation approach effectively preserves code readability while maintaining semantics.
% regarding the different ingredients, we observe that logging level prediction shows little performance decline in 1.1\%, which demonstrates the effectiveness of our code transformation approach in maintaining code readability while preserving semantics.
% Nevertheless, predicting logging variables and logging text generation, which require a comprehensive understanding of the context, experience significant drops in performance of 13.5\% and 15\%, respectively, posing a critical challenge.
% In contrast, it should be noted that the logging variables and logging text generation, which demand a thorough understanding of the context, 
% have been impacted by 13.5\% and 15\%, respectively, and thus pose a critical challenge.

\begin{figure}[tbp]
     \includegraphics[width=\columnwidth]{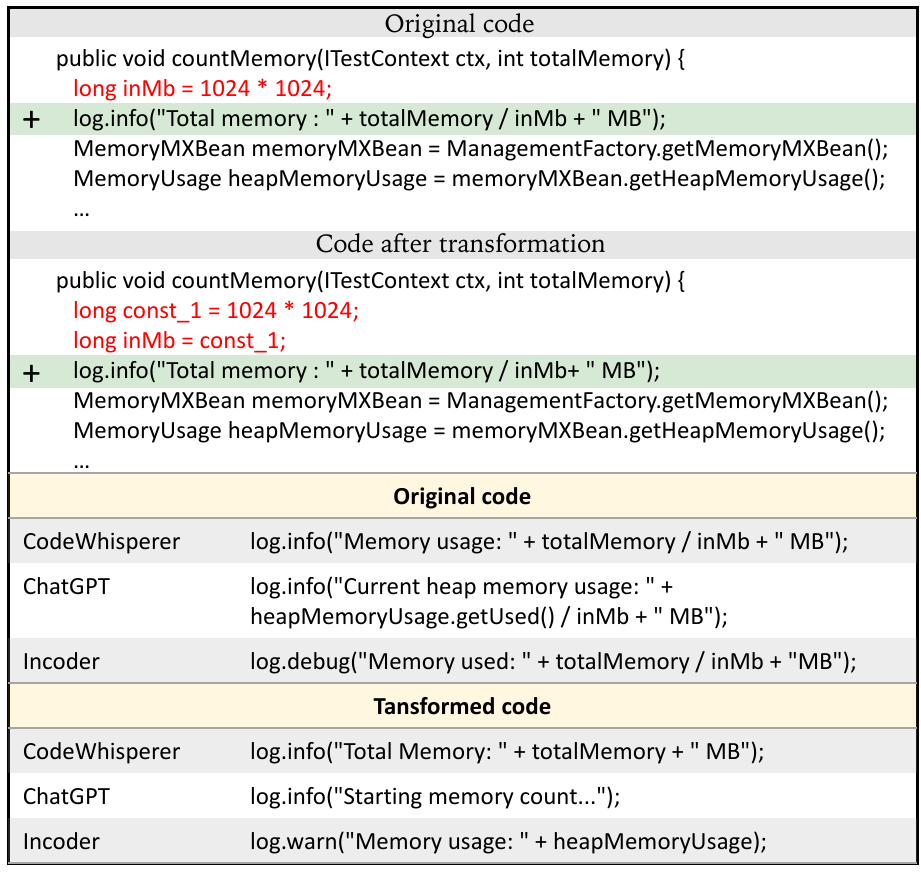}
     \caption{A case of code transformation and its corresponding predicted logging statement from multiple models.}
     \label{fig:RQ4-transformation-case}
 \end{figure}

Fig.~\ref{fig:RQ4-transformation-case} illustrates a transformation case where we highlight code differences in \textcolor{red}{red} and demonstrate how LLMs (CodeWhisperer, ChatGPT, Incoder) log accordingly. 
Regarding the original code, all models correctly predict that \fixedwidth{inMB} should be used to record memory. 
However, after transforming the constant expression \fixedwidth{1024*1024} to a new variable \fixedwidth{const\_1} and then assigning \fixedwidth{const\_1} to \fixedwidth{inMB}, all models fail to understand and identify \fixedwidth{inMB} (or const\_1) as a logging variable. 
CodeWhisperer and Incoder mistakenly predict \fixedwidth{totalMemory} and \fixedwidth{heapMemoryUsage} as the memory size indicator without dividing it by 1024*1024 to be converted into MB units, while ChatGPT does not suggest any variables.
Even though the transformation retains code semantics, existing models exhibit a significant performance drop, indicating their limited generalization abilities.

\begin{tcolorbox}[boxsep=1pt,left=2pt,right=2pt,top=3pt,bottom=2pt,width=\linewidth,colback=white!90!black,boxrule=0pt, colbacktitle=white!,toptitle=2pt,bottomtitle=1pt,opacitybacktitle=0]
\textbf{Finding 8.} \textit{LLMs' performance on variable prediction and logging text generation drops significantly for unseen code by ~\textit{11.6\%} and ~\textit{15.0\%} on average across models, respectively, highlighting the need to improve the generalization capabilities of these models.
}
\end{tcolorbox}

\section{Implications and Advice}\label{sec:discussion}

\textbf{Pay more attention to logging texts.} 
According to Section~\ref{sec:rq1}, while existing models offer satisfactory predictions for logging levels, recommending proper logging variables and logging texts is difficult, particularly the latter.
Since LLMs have shown stronger text generation ability than previous neural networks, future research should focus on using LLMs for the challenging problem of logging text generation instead of simply predicting logging levels.

\begin{tcolorbox}[boxsep=1pt,left=2pt,right=2pt,top=3pt,bottom=2pt,width=\linewidth,colback=white,boxrule=0.5pt, colbacktitle=white!,toptitle=2pt,bottomtitle=1pt,opacitybacktitle=0]
\textbf{Implication 1.} \textit{Future logging studies are encouraged to take advantage of prompting LLMs and focus on the challenging problem of logging text generation.}
\end{tcolorbox}

\textbf{Devise alternative evaluation metrics.} 
Section~\ref{sec:rq1} extensively evaluates the performance of LLMs in generating logging statements using twelve metrics over three ingredients.
We observe that a model may excel in one ingredient while performing poorly in others, and such inconsistency makes any comparison and selection of LLMs difficult.
Existing metrics like BLEU and ROUGE, while suitable and being widely-used~\cite{mastropaolo2022using,ding2022logentext}, may not be optimal for logging statements evaluation because they do not consider semantics when assessing similarity between texts: they aggressively penalize lexical differences, even if the predicted logging statements are synonymous to the actual ones~\cite{wieting2019beyond}.

An alternative perspective to assessing the quality of logging statements involves examining the information entropy for operation engineers. Past research has highlighted that a small number of logging statements often dominate an entire log file~\cite{yu2023logreducer}, posing challenges for engineers in figuring out failure-indicating logs. These limitations underscore the need for a succinct and precise logging strategy in practical applications.

%Therefore, it is interesting to explore an alternative, or even a unified metric integrating all logging statement ingredients, for the purpose of assessing the complete logging statements.

\begin{tcolorbox}[boxsep=1pt,left=2pt,right=2pt,top=3pt,bottom=2pt,width=\linewidth,colback=white,boxrule=0.5pt, colbacktitle=white!,toptitle=2pt,bottomtitle=1pt,opacitybacktitle=0]
\textbf{Implication 2.} \textit{It is recommended to investigate better, possibly unified metrics addressing all ingredients, to evaluate logging statement generation quality.}
\end{tcolorbox}

\textbf{Refine prompts with domain knowledge.}
In Section~\ref{sec:rq3}, we highlight that effective example demonstrations play a crucial role in enhancing the logging performance of LLM by imparting domain knowledge for few-shot learning. Nevertheless, our experiments reveal that augmenting the number of examples does not consistently result in improved performance. 
These insights elicit the development of an advanced selection strategy for choosing demonstrations, aiming to include the most informative ones in the prompt. The selection strategy can draw inspiration from program structure similarity (e.g., try-catch), syntax text similarity (e.g., TF-IDF), or code functional similarity~\cite{DBLP:conf/sigsoft/ZhaoH18}.

% One potential strategy for integrating effective code knowledge is simultaneously fine-tuning LLMs in multiple code tasks, where the shareable code behaviors can be acquired during multi-task learning.
% Taking the tunable model LLAMA~\cite{touvron2023llama} as an example, even though it is trained for general purpose, it can nevertheless continuously incorporate programming knowledge by tuning for multiple tasks, such as comment generation, code type inference, and code infilling.

\begin{tcolorbox}[boxsep=1pt,left=2pt,right=2pt,top=3pt,bottom=2pt,width=\linewidth,colback=white,boxrule=0.5pt, colbacktitle=white!,toptitle=2pt,bottomtitle=1pt,opacitybacktitle=0]
\textbf{Implication 3.} \textit{Designing a demonstration selection framework for effective few-shot learning can yield better results.}
\end{tcolorbox}

\textbf{Provide broader programming contexts for LLMs.}
In Section~\ref{sec:rq4}, we investigate how expanding programming contexts can significantly enhance the logging performance of LLMs. Such a finding implies that extending the context to the file level, rather than the method level, is beneficial for acquiring extra information as well as learning logging styles.
However, including the entire repository as input for LLMs may be impractical for large programs due to input token limitations. Additionally, LLM performance tends to decline with longer inputs, even when within the specified context length~\cite {shi2023large, liu2024lost}. To capture effective programming contexts for specific methods, a promising solution involves identifying methods with associated calling relationships and variable definitions. Providing methods spanning multiple classes can also contribute to generating logging statements consistent with existing ones, thereby learning intra-project logging styles.

% However, it is impractical to directly feed the whole repository into LLMs due to the input limits and 
% In addition, broadening the programming contexts to the file level may further allow LLMs to uncover variable definitions and learn intra-project logging styles over multiple classes, which help to generate logging statements that are consistent with existing ones.

\begin{tcolorbox}[boxsep=1pt,left=2pt,right=2pt,top=3pt,bottom=2pt,width=\linewidth,colback=white,boxrule=0.5pt, colbacktitle=white!,toptitle=2pt,bottomtitle=1pt,opacitybacktitle=0]
\textbf{Implication 4.} \textit{When using LLMs for logging, future research could broaden the programming context by incorporating information from function invocations and variable definitions.}
\end{tcolorbox}

\textbf{Enhance generalization capabilities of LLMs.}
In Section~\ref{sec:rq4}, we observe that current LLMs show significantly worse performance on unseen code, reflecting their limited generalization capabilities. 
The result can be attributed to the capacity of parameters in LLMs to memorize large datasets~\cite{rabin2023memorization}.
% models with excessive parameter capacity easily memorize large datasets
This issue will become more severe when tackling code in a rapidly evolving software environment, resulting in more unseen code.
One effective idea is to apply a prompt-based method with few chain-of-thought demonstrations~\cite{rubin2021learning, wei2022chain} to foster the generalization capabilities of ever-growing LLMs. The chain-of-thought strategy allows models to decompose complicated multi-step problems into several intermediate reasoning steps. For example, we can ask models to focus on special code structures (e.g., if-else), then advise them to elicit key variables and system activities to log. While the chain-of-thought strategy has shown success in natural language reasoning tasks~\cite{kojima2022large}, 
future work should explore such prompt-based approaches to enhance generalization capabilities.

% the chain-of-thought strategy~\cite{wei2022chain}, which is a prompt-based method with few chain-of-thought demonstrations provided. 

\begin{tcolorbox}[boxsep=1pt,left=2pt,right=2pt,top=3pt,bottom=2pt,width=\linewidth,colback=white,boxrule=0.5pt, colbacktitle=white!,toptitle=2pt,bottomtitle=1pt,opacitybacktitle=0]
\textbf{Implication 5.} \textit{We should investigate prompt-based strategies with zero-shot or few-shot learning to improve the generalization ability of LLMs.}
\end{tcolorbox}

\section{Threats to Validity}\label{sec:threats}
%\yc{2} Besides, the ability of ROUGE and BLEU scores to accurately measure logging text has long a subject of discussion~\cite{mastropaolo2022using,ding2022logentext} same as we discussed in Sec~\ref{sec:discussion}. Despite the possible lack of precision in the evaluation of logging text metrics, the scores of logging levels, logging variables, and logging text exhibit a consistent trend in evaluating the logging statement generation capabilities of these LLMs. This consistency suggests that it does not hinder the effectiveness of our evaluation, and we further elaborate on our understanding of the unified metric in Sec~\ref{sec:discussion}.
\noindent \textbf{Internal Threats.} 
(1) A concern of this study is the potential bias introduced by the limited size of the \Odata~dataset, which consists of \textit{3,840} methods.
This limitation arises due to the fact that those plugin-based code completion tools impose usage restrictions to prevent bots; therefore, human efforts are needed.
To address the threat, we acquired and sampled \Odata~and \Tdata~datasets from well-maintained open projects, which we believe are representative.
% maintain that both the \Odata~and \Tdata~datasets are adequately representative for evaluation since they are derived and sampled from well-maintained open projects and automatically transformed to prevent data leakage issues. 
Note that existing Copilot testing studies also have used datasets of comparable sizes~\cite{mastropaolo2023robustness,pearce2022asleep}.

(2) Another concern involves the context length limitations of certain language models~\cite{fried2022incoder, ChatGPT, gpt-3.5} (e.g., \textit{4,097} tokens for Davinci), which may affect the file-level experiment. 
To address this concern, we analyze the collected data and reveal that
% performed an analysis of the data we collected, focusing on the size of Java files in terms of tokens. Our analysis revealed that 
\textit{98.6\%} of the Java files fall within the \textit{4096}-token limit, and \textit{94.3\%} of them are within the \textit{2048}-token range. Such analysis implies that the majority of files in our dataset remain unaffected by the context length restrictions.
% Consequently, we argue that the context length limitations of these models do not significantly compromise the validity of our experiment. 

(3) The other threat is the potential effect of various prompts on Davinci and ChatGPT. To address this, we invited four authors to independently provide three prompts according to their usage habits. These prompts were evaluated using a dataset of 100 samples, and the one that demonstrated the best performance was selected. This approach ensures that the chosen prompt is representative for daily development.
% This approach ensures that the chosen prompt is not only effective but also representative of actual usage scenarios in daily development.

\noindent \textbf{External Threats.} One potential external threat stems from the fact that the~\Odata~dataset was mainly based on the Java language, which may affect the generalizability of our findings to other languages. 
However, according to previous works~\cite{li2021deeplv, liu2022tell, mastropaolo2022using}, Java is among the most prevalent programming languages for logging research purposes, and both SLF4J and Log4j are highly popular and widely adopted logging APIs within the Java ecosystem. We believe the representativeness of our study is highlighted by the dominance of Java languages and these APIs in the logging domain. The core idea of the study can still be generalized to other logging frameworks or languages.

\section{Related Work}\label{sec:relatedwork}

\subsection{Logging Statement Automation}

The logging statement automation studies focus on automatically generating logging statements, which can be divided into two categories: \textit{what-to-log} and \textit{where-to-log}.
\textit{What-to-log} studies are interested in producing concrete logging statements, which include deciding the appropriate log level (e.g., warn, error)~\cite{li2021deeplv, liu2022tell, li2017log}, choosing suitable variables~\cite{liu2019variables, dai2022reval, yuan2012improving}, and generating proper logging text~\cite{mastropaolo2022using, ding2022logentext}. For example, ordinal-based neural networks~\cite{li2021deeplv} and graph neural networks~\cite{liu2022tell} have been applied to learn syntactic code features and semantic text features for log-level suggestions. LogEnhancer~\cite{yuan2012improving} aims to reduce the burden of failure diagnosis by inserting causally-related variables in a logging statement from a programming analysis perspective, whereas Liu et al.~\cite{liu2019variables} predicts logging variables for developers using a self-attention neural network to learn tokens in code snippets. 
%Besides,~\citet{mastropaolo2022using} adapts from a Text-To-Text-Transfer-Transformer (T5)~\cite{raffel2020exploring} model to automate logging activities (including generating logging statement text) after pretraining it on a large corpus.
\textit{Where-to-log} studies concentrating on suggesting logging points in source code~\cite{li2020shall, zhao2017log20}. Excessive logging statements can enhance unnecessary efforts in software development and maintenance, while insufficient logging statements lead to missing key system behavior information for potential system diagnosis~\cite{fu2014developers, zhu2015learning}. To automate logging points, previous studies solve the log placement problem in specific code construct types, such as catch~\cite{lal2016logoptplus}, if~\cite{lal2016logoptplus}, and exception~\cite{yuan2012conservative}. Li et al.~\cite{li2020shall} proposes a deep learning-based framework to suggest logging locations by fusing syntactic, semantic and block features extracted from source code. 
The most recent model in T5 architecture, LANCE~\cite{mastropaolo2022using}, provides a one-stop logging statements solution for deciding logging points and logging contents for code snippets.

Although these works tried new emerging deep-learning models to determine logging statements, they have certain limitations: some focus solely on specific logging ingredients or are designed for particular scenarios.
Consequently, these work and their proposed datasets, holding different experimental settings, which are not well-suited for evaluating logging ability for daily development.
Moreover, they lack the analysis of the model itself (e.g., potential influencing factors) and comprehensive evaluation (e.g., performance across multiple ingredients).
To fill the gap, our study is the first one that investigates and compares current LLMs for automated logging generation, which facilitates future research in developing, applying, and integrating these large models in practice.

% Moreover, they lack an analysis of the model itself, such as the potential influencing factors, and
% Moreover, they lack the analysis of the model itself and comprehensive evaluation, for example, how to measure the generated logging statements, or in what scenario can the model performs better.

\subsection{Empirical Study on Logging Practice}
Logging practices have been widely studied to guide developers in writing appropriate logging statements, because modern log-based software maintenance highly depends on the quality of logging code~\cite{yuan2012improving,chen2021survey, ding2015log2}. Logging too little or too much will both hinder the failure diagnosis process~\cite{chen2017characterizing}.
To reveal how logging practices in the industry help engineers make logging decisions, Fu et al.~\cite{fu2014developers} analyzes two large-scale online service systems involving 54 experienced developers at Microsoft, providing six insightful findings concerning the logging code categories, decisional factors, and auto-logging feasibility. Another industrial study~\cite{pecchia2015industry} indicates that the logging process is developer-dependent and thus strongly suggested standardizing event logging activities company-wide.
Exploration studies on logging statements' evolution over open software projects have also been conducted~\cite{chen2017characterizing, kabinna2018examining, shang2014exploratory}, revealing that paraphrasing, inserting, and deleting logging statement operations are prevalent during software evolution. 
Chen et al.~\cite{chen2021survey} revisits the logging instrumentation pipeline with three phases, including logging approach, logging utility integration, and logging code composition. 
While some studies~\cite{he2021survey,chen2021survey} introduce the existing what-to-log approaches with technical details, their main emphasis lies in the overall log workflow, encompassing proactive logging generation and reactive log management.
However, they do not offer a qualitative comparison or a discussion on the characteristics of the logging generation tools.

In summary, even though logging practices have been widely studied as a crucial part of software development, there exists neither a benchmark evaluation of logging generation models nor a detailed analysis of them.
To bridge the gap, this study is the first empirical analysis of LLM-based logging statement generation tools by benchmarking existing solutions. The findings and implications can further guide researchers to build more effective and practical automated logging models.

\subsection{Large Language Models for Code}
The remarkable success of LLMs in the NLP has prompted the development of pre-trained models in other areas, particularly in intelligent code analysis~\cite{xia2023automated, copilot_doc,clement2020pymt5}.
CodeBERT~\cite{feng2020codebert} adopts the transformer architecture~\cite{vaswani2017attention} and has been trained on a blend of programming and natural languages to learn a general representation for code, which can further support generating a program from a natural language specification.
In addition to sequence-based models, GraphCodeBERT~\cite{guo2020graphcodebert} considers the code property of
its structural and logical relationship (e.g., data flow, control flow), creating a more effective model for code understanding tasks~\cite{karmakar2021pre}.
Furthermore, Guo et al.~\cite{guo2022unixcoder} presents UniXCoder, which is a unified cross-modal pre-trained model for programming language. UniXcoder employs a mask attention mechanism to regulate the model's behavior and trains with cross-modal contents such as AST and code comment to enhance code representation.
The recent work, InCoder~\cite{fried2022incoder}, is adept at handling generative tasks (e.g., comment generation) after learning bidirectional context for infilling arbitrary code lines.

% With the increasing implementation of large code models, many code models have also been integrated into IDE plugins~\cite{aiXcoder, copilot_doc, codegeex, tabnine}, effectively assisting in daily development tasks. 
As the use of large code models grows, many of them have been integrated into IDE plugins~\cite{codegeex, tabnine, copilot_doc, aiXcoder} to assist developers in their daily programming. 
Nonetheless, existing code intelligence research focuses on functional code and these non-functional logging statements have never been explored. By extensively examining the performance of LLMs in writing logging statements, this paper contributes to a deeper understanding of the potential applications of LLMs in automated logging.

%Numerous efforts~\cite{copilot_doc, xu2022codeparrot, black2021gpt, guo2022unixcoder, fried2022incoder, codegeex, feng2020codebert, wang2021codet5, ChatGPT} have been dedicated to advancing automatic code completion by training large models on code tokens. Neural code completion models often benefit from large-scale pretraining on massive codebases, which enables them to learn general programming patterns and language-specific syntax. Fine-tuning on smaller, task-specific datasets helps tailor these pre-trained models to specific domains or coding styles~\cite{wang2021codet5, feng2020codebert, fried2022incoder}. Furthermore, by incorporating surrounding comments or direct instructions (e.g., ChatGPT~\cite{ChatGPT}) as prompts, code completion models~\cite{ChatGPT, codegeex, copilot_doc, guo2022unixcoder} can generate more precise code snippets, resulting in improved performance and accuracy~\cite{wang2022no}, thereby delivering a more tailored and effective coding experience for developers.

% To our best knowledge, this empirical study serves as the first empirical study on logging statement generation tools.

% This left a blank for developers to measure whether they can use automated logging assistants when writing code.
% Modern LLMs enable the solutions to this challenging yet critical logging statement generation problem.
\section{Conclusion} \label{sec:conclusion}

In this paper, we present the first extensive evaluation of LLMs for generating logging statements. 
To achieve this, we introduce a logging statement generation benchmark dataset, LogBench, and assess the effectiveness and generalization capabilities of eleven top-performing LLMs.
While LLMs are promising in generating complete logging statements, they can still be promoted in multiple ways.

First, our evaluation indicates that existing LLMs are not yet adept at generating complete logging statements, particularly in producing effective logging texts. Nonetheless, their direct application surpasses the performance of conventional logging models, indicating a promising future for leveraging LLMs in logging practices.

In addition, we delve into the construction of prompts that influence LLMs' logging performance, considering factors such as instructions and the number of example demonstrations. While our experiments demonstrate the advantages of incorporating demonstrations, we observe that an increased number of demonstrations does not consistently result in improved logging performance. Thus, we recommend the development of a demonstration selection framework in future research. Furthermore, we identify external factors, such as comments and programming contexts, that enhance model performance. We encourage the incorporation of such factors to enhance LLM-based logging tools.

Last but not least, we evaluate LLMs' generalization ability using a dataset that includes transformed code. 
Our findings indicate that directly applying LLMs to unseen code results in a significant decline in performance, highlighting the necessity to enhance their inference abilities.
We suggest employing the chain-of-thought technologies to break down the logging task into smaller logical steps as a future step, unlocking LLMs' full potential.
% Lastly, we derive five implications for future research on adopting language models for automated logging generation.
We hope this paper can stimulate more work in the promising direction of using LLMs for automatic logging.

\section{Data Availability} \label{sec:availability}

% All collected data of our study \Odata and \Tdata, source code and code transformation tool we developed are available at the anonymous Github link \url{https://github.com/LoggingResearch/LoggingStudy}.

The datasets \Odata and \Tdata, source code, and code transformation tool are available at the anonymous Github link:  \url{https://github.com/LoggingResearch/LoggingStudy}.

\balance
\bibliographystyle{IEEEtran}
\bibliography{tse-base}

\end{document}